\documentclass[journal,onecolumn]{IEEEtran}

\usepackage{etex}
\newcommand{\subparagraph}{}
\usepackage{hyperref}
\usepackage{amsmath}
\usepackage{amsthm}
\usepackage{bbm}
\usepackage[utf8]{inputenc}
\usepackage{graphicx}
\usepackage{caption}
\usepackage{subcaption}
\usepackage{framed, color} 
\usepackage[margin=.7in]{geometry}
\usepackage{pdflscape}
\definecolor{shadecolor}{rgb}{1, 0.8, 0.3}
\usepackage{amssymb}
\usepackage{amsfonts}
\usepackage{cite}
\usepackage{ifthen}
\usepackage{epsfig}
\usepackage{subcaption}
\usepackage{array}
\newcolumntype{C}[1]{>{\centering\let\newline\\\arraybackslash\hspace{0pt}}m{#1}}
\usepackage{enumerate}
\usepackage{algorithm}
\usepackage[all]{xy}
\usepackage{algpseudocode}
\usepackage[compact]{titlesec}
\usepackage{footmisc}
\usepackage{times}
\usepackage{tikz}
\usepackage{float}
\usetikzlibrary{shapes,positioning,arrows,automata}
\usepackage{caption}
\usepackage{nth}
\usepackage{array} 
\theoremstyle{plain}
\newtheorem{theorem}{Theorem}

\newtheorem{fact}{Fact}
\newtheorem{lemma}{Lemma}

\newtheorem{corollary}{Corollary}

\theoremstyle{definition}

\theoremstyle{remark}

\usepackage{array}

\newcommand{\kjp}{P}
\newcommand{\pp}{the parity measurement}
\newcommand{\hh}{the homogeneity measurement}
\newcommand{\rr}{random LDGM codes with a constant right-degree $d$}

\newcommand{\beq}{\begin{equation}}
	\newcommand{\eeq}{\end{equation}}
\newcommand{\bea}{\begin{eqnarray}}
	\newcommand{\eea}{\end{eqnarray}}
\newcommand{\bean}{\begin{eqnarray*}}
	\newcommand{\eean}{\end{eqnarray*}}
\newcommand{\bit}{\begin{itemize}}
	\newcommand{\eit}{\end{itemize}}
\newcommand{\ben}{\begin{enumerate}}
	\newcommand{\een}{\end{enumerate}}
\newcommand{\blem}{\begin{lem}}
	\newcommand{\elem}{\end{lem}}
\newcommand{\bthm}{\begin{thm}}
	\newcommand{\ethm}{\end{thm}}
\newcommand{\bpf}{\begin{IEEEproof}}
	\newcommand{\epf}{\end{IEEEproof}}

\newcommand{\comment}[1]{}

\usepackage{algorithm}\usepackage{algpseudocode}
\usepackage{mathtools}
\usepackage{soul}

\newcommand{\ye}{f_E}
\newcommand{\Ye}{\mathbf{F}}

\newcommand{\ex}{\mathbb{E}}

\usepackage{epstopdf}

\def\SingleColumn{1}
\titleformat{\subsubsection}[runin]{\normalfont}{\arabic{subsubsection})}{0.5ex}{}[:]

\begin{document}

\title{}

\author{\IEEEauthorblockN{Kwangjun Ahn}
\IEEEauthorblockA{Dept. of Mathematical Sciences\\
KAIST\\
kjahnkorea@kaist.ac.kr}
\and
\IEEEauthorblockN{Kangwook Lee}
\IEEEauthorblockA{School of EE\\KAIST\\
kw1jjang@kaist.ac.kr}
\and
\IEEEauthorblockN{Changho Suh}
\IEEEauthorblockA{School of EE\\KAIST\\
chsuh@kaist.ac.kr}
}

\title{Community Recovery in Hypergraphs}
\author{
	Kwangjun Ahn$^*$, Kangwook Lee$^*$, and Changho Suh
	\thanks{
		
		Kwangjun Ahn is with the Department of Mathematical Sciences, KAIST (e-mail: kjahnkorea@kaist.ac.kr).		 
		Kangwook Lee and Changho Suh are with the School of Electrical Engineering, KAIST (e-mail: $\{$kw1jjang, chsuh$\}$@kaist.ac.kr).

		This paper was presented in part at the 54th Annual Allerton Conference on Communication, Control, and Computing 2016~\cite{ahn2016community}, and the IEEE International Symposium on Information Theory 2017~\cite{Ahn1706:Information}.
		
		$*$ Kwangjun Ahn and Kangwook Lee contributed equally to this work
		}}

\date{}
\maketitle

\begin{abstract}

Community recovery is a central problem that arises in a wide variety of applications such as network clustering, motion segmentation, face clustering and protein complex detection. 
The objective of the problem is to cluster data points into distinct communities based on a set of measurements, each of which is associated with the values of a certain number of data points. 
While most of the prior works focus on a setting in which the number of data points involved in a measurement is two, this work explores a generalized setting in which the number can be more than two. 
Motivated by applications particularly in machine learning and channel coding, we consider two types of measurements: (1) \emph{homogeneity} measurement which indicates whether or not the associated data points belong to the same community; (2) \emph{parity} measurement which denotes the modulo-2 sum of the values of the data points.
Such measurements are possibly corrupted by Bernoulli noise.
We characterize the fundamental limits on the number of measurements required to reconstruct the communities for the considered models.

\end{abstract}

\section{Introduction}
Clustering of data is one of the central problems, and it arises in many fields of science and engineering.
Among many related problems, \emph{community recovery in graphs} has received considerable attention with applications in numerous domains such as social networks~\cite{girvan2002community, fortunato2010community,porter2009communities}, computational biology~\cite{chen2006detecting}, and machine learning~\cite{huang2013consistent,shi2000normalized}. 
The goal of the problem is to cluster data points into different communities based on \emph{pairwise} information. Among a variety of models for the community recovery problem,
the stochastic block model (SBM)~\cite{holland1983stochastic} and the censored block model (CBM)~\cite{abbe2013conditional} have received significant attention in recent years.
In SBM, two data points in the same communities are more likely to be connected by an edge than the other edges. 
In the case of CBM, each measurement returns the modulo-$2$ sum of the values assigned to the two nodes, possibly corrupted by Bernoulli noise.

While these models reflect interactions between a pair of two nodes, there are numerous applications in which interactions occur across more than two nodes.
One such application is a folksonomy, a social network in which users can annotate items with different tags~\cite{ghoshal2009random}.
In this application, the graph consists of nodes corresponding to different users, different items, and different tags. 
When user $i$ annotate item $j$ with tag $k$, one can view this as a hyperedge connecting node $i$, node $j$ and node $k$.
Therefore, in order to cluster nodes of such a graph based on such interactions, one needs a model that can capture such three-way interactions. 
Another application is molecular biology, in which multi-way interactions between distinct systems capture complex molecular interactions~\cite{michoel2012alignment}.
There are also a broad range of applications in other domains including computer vision~\cite{agarwal2006higher}, VLSI circuits~\cite{karypis2000multilevel}, and categorical databases~\cite{gibson1998clustering}.

These applications naturally motivate us to investigate a \emph{hypergraph} setting in which measurements are of \emph{multi-way} information type.
Specifically, we consider a simple yet practically-relevant model, which we name the generalized censored block model (GCBM).
In the GCBM, the $n$ data points are modeled as nodes in a \emph{hypergraph}, and their interactions are encoded as hyperedges between the nodes. 
As an initial effort, we focus on a simple setting in which there are two communities: each node taking either 0 or 1 depending on its affiliation.
More concretely, we consider a random $d$-uniform hypergraph in which each hyperedge connecting a set of $d$ nodes exists with probability $p$ and takes a function of the values assigned to the $d$ nodes.
In this work, inspired by applications in machine learning and channel coding, we study the following two types of measurements:
\begin{itemize}
	\item  \emph{\hh} that reveals whether or not the $d$ nodes are in the same community; and
	\item \emph{\pp} that reveals the modulo-$2$ sum of the affiliation of the $d$ nodes.
\end{itemize}
Further, we study both the noiseless case and the noisy case.

\subsection{Main contributions}
Specialized to the $d=2$ case, the above two measurement models reduce to the CBM, in which the information-theoretic limit on the expected number of edges required for exact recovery is characterized as $p \binom{n}{2}=\frac{1}{2}\cdot\frac{n \log n}{\left(\sqrt{1-\theta}-\sqrt{\theta}\right)^2}$~\cite{abbe2014decoding, 7523889}. 
On the other hand, the information-theoretic limits for the case of arbitrary $d$ has not been settled.
This precisely sets the goal of our paper:
We seek to characterize the information-theoretic limits on the sample complexity for exact recovery under the two models. 
A summary of our findings is as follows. For a fixed constant $d$, the information-theoretic limits are: 
\begin{itemize}
	\item  (\hh~case) $p {n \choose d} =\frac{2^{d-2}}{d}\cdot\frac{n \log n}{\left(\sqrt{1-\theta}-\sqrt{\theta}\right)^2}$ if $d$ is a fixed constant; and
	\item (\pp~case) $p {n \choose d} =\frac{1}{d}\cdot\frac{n \log n}{\left(\sqrt{1-\theta}-\sqrt{\theta}\right)^2}$ if $d$ is a fixed constant.
\end{itemize}
For \pp~case, we also characterize the information-theoretic limits for a more general setting where $d$ can arbitrarily scale with $n$. 
\begin{itemize}
	\item (\pp~case) $p {n \choose d} =\Theta \left(\frac{n \log n}{d}\right)$ if $d = o(\log n)$; and
	\item (\pp~case) $p {n \choose d} = \Theta( n)$ if $d = \Omega(\log n)$.
\end{itemize}

\begin{table*}[t]
	\caption{{\bf Summary of main results.} The information-theoretic limits on sample complexity ($p\binom{n}{d}$) are summarized. Here, $n$ denotes the number of nodes, $\theta$ denotes the noise probability, and $d$ denotes the size of hyperedges. ``$d=f(n)$'' means that $d$ can scale with $n$, and ``$\Theta_{\theta,d}$'' implies that the constant involved depends on $\theta$ and $d$.}
	\label{table1}
	\begin{center}
		\begin{tabular}{| c || c | c| c| }
			\hline		
			  &$d=2$ &$d>2$ (const.) & $d=f(n)$\\
			 \hline \hline
			Homogeneity & $\frac{1}{2}\cdot\frac{n \log n}{\left(\sqrt{1-\theta}-\sqrt{\theta}\right)^2}$ & $\frac{2^{d-2}}{d}\cdot\frac{n \log n}{\left(\sqrt{1-\theta}-\sqrt{\theta}\right)^2}$ & N/A\\
			Parity &$\frac{1}{2}\cdot\frac{n \log n}{\left(\sqrt{1-\theta}-\sqrt{\theta}\right)^2}$ & $\frac{1}{d}\cdot \frac{n \log n}{\left(\sqrt{1-\theta}-\sqrt{\theta}\right)^2}$ & $\Theta_{\theta,d}\left( \max\left\{n,~ \frac{n\log n}{d}\right\}\right)$\\
			\hline  		
		\end{tabular}
	\end{center}
\end{table*}

These results provide some interesting implications to relevant applications such as subspace clustering and channel coding. 
In particular, the results offer concrete guidelines as to how to choose $d$ that minimizes sample complexity while ensuring successful clustering. 
See details in Sec.~\ref{sec:model_just} and Sec.~\ref{sec:MainResults}.

\subsection{Related work}

\subsubsection{The $d=2$ case} 
The exact recovery problem in standard graphs ($d=2$) has been studied in great generality. 
In SBM, both the fundamental limits and computationally efficient algorithms are investigated initially for the case of two communities~\cite{abbe2016exact,7523889,MNS14a}, and recently for the case of an arbitrary number of communities~\cite{abbe2015community}.
In CBM, \cite{abbe2014decoding} characterizes the sample complexity limit, and \cite{7523889} develops a computationally efficient algorithm that achieves the limit.

Another important recovery requirement is \emph{detection}, which asks whether one can recover the clusters better than a random guess. The modern study of the detection problem in SBM is initiated by a paper by Decelle et al.~\cite{decelle}, which conjectures
phase transition phenomena for the detection problem\footnote{In the paper, it is also conjectured that an information-computation gap might exist for the case of more than $3$ communites ($k\geq 4$). This conjecture is also extensively studied in~\cite{YC14, NN14, Mon15,BM16}, and is recently settled in~\cite{abbe2015detection}.}.
This conjecture is initially tackled for the case of two communities. 
The impossibility of the detection below the conjectured threshold is established in~\cite{mossel2015reconstruction}, and it is proved in~\cite{MNS14b, Mas14,bordenave} that the conjectured threshold can be achieved efficiently.
The conjecture for the arbitrary number of communities is recently settled by Abbe and Sandon~\cite{abbe2015detection}.
For another line of researches, minimax-optimal rates are derived in~\cite{zhang2016minimax}, and algorithms that achieve the rates are developed in~\cite{gao2015achieving}. 
We refer to a recent survey by Abbe~\cite{abbe2017community} for more exhaustive information.

\subsubsection{The homogeneity measurement case}
 Recently, \cite{JMLR:v18:16-100,ghoshdastidar2015consistency} consider a general model that includes our model as a special case (to be detailed in Sec.~\ref{sec:PF}), and provide an upper bound on sample complexity for \emph{almost exact} recovery, which allows a vanishing fraction of misclassified nodes.
Applying their results to our model, their upper bound reduces to $p {n \choose d}= \Omega(n \log^2n)$.
Whether or not the sufficient condition is also necessary has been unknown.
 In this work, we show that it is not the case, demonstrating that the minimal sample complexity even for exact recovery is $\Theta (n \log n)$.
  
We note that the homogeneity measurement case is closely related to subspace clustering, one of the popular problems in computer vision~\cite{govindu2005tensor, chen2009spectral,agarwal2006higher}; See Sec.~\ref{sec:subspace} for details.

\subsubsection{The parity measurement case} 
The parity measurement case has been explored by~\cite{watanabe2013message} in the context of random constraint satisfaction problems.
The case of $d=3$ has been well-studied: it is shown that the maximum likelihood decoder succeeds if $p{n \choose 3} \geq 2\cdot \frac{n\log n}{(0.5-\theta)^2}$~\cite{watanabe2013message}. 
Unlike the prior result which only considers the case of $d=3$, we cover an arbitrary constant $d$, and characterize the sharp threshold on the sample complexity.  

Abbe-Montanari~\cite{abbe2013conditional} relate the parity measurement model to a channel coding problem in which \rr~are employed.
By proving the concentration phenomenon of the mutual information between channel input and output, they demonstrate the existence of phase transition for an even $d$.
Our results span \emph{any} fixed $d$, and hence fully settle the phase transition (see Sec.~\ref{sec:MainResults}).

\subsubsection{The stochastic block model for
	hypergraphs}
There are several works which study the community recovery under SBM for hypergraphs.
In~\cite{florescu2015spectral},  the authors explore the case of two equal-sized communities\footnote{Actually, the main model in the paper is \emph{the bipartite stochastic block model}, which is not a hypergraph model. However, the result for the hypergraph case follows as a corollary (see Theorem 5 therein).}. Specializing it to our model, one can readily show that detection is possible if $\binom{n}{d}p =\Omega(n)$. 
Moreover, \cite{angelini} recently conjectures phase transition thresholds for detection.
Lastly, \cite{wangetal} derives the minimax-optimal error rates, and generalizes the results in \cite{zhang2016minimax} to the hypergraph case.

\subsubsection{Other relevant problems}
Community recovery in hypergraphs bears similarities to other inference problems, in which the goal is to reconstruct data from multiple queries. Those problems include crowdsourced clustering~\cite{vesdapunt2014crowdsourcing,ashtiani2016clustering}, group testing~\cite{dorfman1943detection} and data exactration from histogram-type information~\cite{7852208,7541526}. Here, one can make a connection to our problem by viewing each query as a hyperedge measurement. 
However, a distinction lies in the way that queries are collected. For instance, an adaptive measurement model is considered in the crowdsourced setting~\cite{vesdapunt2014crowdsourcing,ashtiani2016clustering} unlike our non-adaptive setting in which hyperedges are sampled uniformly at random. 
Histogram-type information acts as a query in~\cite{dorfman1943detection,7852208,7541526}.

\subsection{Paper organization}
Sec.~\ref{sec:PF} introduces the considered model; 
in Sec.~\ref{sec:MainResults}, our main results are presented along with some implications;
in Sec.~\ref{pf:thm1},~\ref{pf:thm2} and~\ref{pf:thm3}, we provide the proofs of the main theorems;
Sec.~\ref{sec:simulation} presents experimental results that corroborate our theoretical findings and discuss interesting aspects in view of applications;
and in Sec.~\ref{sec:conclusion}, we conclude the paper with some future research directions. 

\subsection{Notations}
For any two sequences $f(n)$ and $g(n)$: $f(n) = \Omega(g(n))$ if there exists a positive constant $c$ such that $f(n)\geq cg(n)$;
 $f(n)=O(g(n))$ if there exists a positive constant $c$ such that $f(n)\leq c g(n)$;
$f(n) = \omega(g(n))$ if $\lim_{n\rightarrow \infty} \frac{f(n)}{g(n)} =\infty$; $f(n) = o(g(n))$ if $\lim_{n\rightarrow \infty} \frac{f(n)}{g(n)} =0$;
 and $f(n)\asymp g(n)$ or $f(n)=\Theta(g(n))$ if there exist positive constants $c_1$ and $c_2$ such that $c_1g(n)\leq f(n)\leq c_2g(n)$.

 For a set $A$ and an integer $m\leq |A|$, we denote $\binom{A}{m} := \{B \subset A \,:\, |B|=m \}.$ 
 Let $[n]$ denote $\{1,\cdots,n\}$.
  Let $\mathbf{e}_i$  be the $i^{\text{th}}$ standard unit vector. 
  Let $\mathbf{0}$ be the all-zero-vector and $\mathbf{1}$ be the all-one-vector. 
 We use $\mathbb{I}\{\cdot\}$ to denote an indicator function.
 Let ${\sf D_{KL}}(p\|q)$ be the Kullback-Leibler (KL) divergence between ${\sf Bern}(p)$ and ${\sf Bern}(q)$, i.e., ${\sf D_{KL}}(p\|q) := p\log \frac{p}{q}+(1-p)\log\frac{1-p}{1-q}$.
 We shall use $\log(\cdot)$ to indicate the natural logarithm.
 We use ${H}(\cdot)$ to denote the binary entropy function.

\section{Generalized censored block models}
\label{sec:PF}

 Consider a collection of $n$ nodes $\mathcal{V} = [n]$, each represented by a binary variable $X_i\in\{0,\,1\}$, $1\leq i \leq n$.  
Let $\mathbf{X}:=\{X_i\}_{1\leq i \leq n}$ be the ground-truth vector.
Let $d$ denote the size of a hyperedge. 
 Samples are obtained as per a \emph{measurement hypergraph} $\mathcal{H} = (\mathcal{V},\mathcal{E})$ where $\mathcal{E}\subset \binom{[n]}{d}$.
We assume that each element in $\binom{[n]}{d}$ belongs to $\mathcal{E}$ independently with probability $p\in[0,1]$. \emph{Sample complexity} is defined as the number of hyperedges in a random measurement hypergraph, which is concentrated around  $p\binom{n}{d}$ in the limit of $n$.
Each sampled edge $E\in \mathcal{E}$ is associated with a noisy binary measurement $Y_E$:
\begin{align}
Y_{E} = f(X_{i_1},X_{i_2},\cdots, X_{i_d}) \oplus Z_E, \label{def:model}
\end{align} where $f: \{0,1\}^d \to \{0,1\}$ is some binary-valued function, $\oplus$ denotes modulo-2 sum, and $Z_E \overset{\text{i.i.d.}}{\sim} {\sf Bern}(\theta)$ is a random variable with noise rate $0\leq \theta<\frac{1}{2}$. 
For the choice of $f$, we focus on the two cases:
\begin{itemize}
\item \emph{\hh:} 
\begin{align*}
~~~\,f_h(X_{i_1},X_{i_2},\cdots, X_{i_d}) = \mathbb{I} \{X_{i_1}= X_{i_2}=\cdots= X_{i_d} \}; 
\end{align*}
\item \emph{ \pp:} 
	\begin{align*}
	f_p(X_{i_1},X_{i_2},\cdots, X_{i_d}) = X_{i_1}\oplus X_{i_2}\oplus \cdots \oplus X_{i_d}.
	\end{align*}
	
\end{itemize} 
Let $\mathbf{Y}:= \{Y_{E}\}_{E\in\mathcal{E}}.$
We remark that when $d=2$, this reduces to CBM~\cite{abbe2014decoding}. 

The goal of this problem is to recover $\mathbf{X}$ from $\mathbf{Y}$. 
In this work, we will focus on the case of even $d$ since the case of odd $d$ readily follows from the even case~\cite{ahn2016community}.
When $d$ is even, the conditional distribution of $\mathbf{Y}|\mathbf{X}$ is equal to that of $\mathbf{Y}|\mathbf{X}\oplus\mathbf{1}$.
Hence, given a recovery scheme $\psi$, the probability of error is defined as 
\[
P_e(\psi) := \max_{\mathbf{X}\in \{0,1\}^n}  \Pr\left(\psi(\mathbf{Y}) \notin \{\mathbf{X},~ \mathbf{X}\oplus \mathbf{1}\}\right).
\]
We intend to characterize the minimum sample complexity, above which there exists a recovery algorithm $\psi$ such that 
$P_e(\psi) \to 0$ as $n$ tends to infinity, and under which $P_e(\psi) \nrightarrow 0$ for all algorithms. 

\subsection{Relevant applications}\label{sec:model_just}

\subsubsection{Subspace clustering and the homogeneity measurement} \label{sec:subspace}
	\begin{figure}
		\centering
		\includegraphics[width=.45\textwidth]{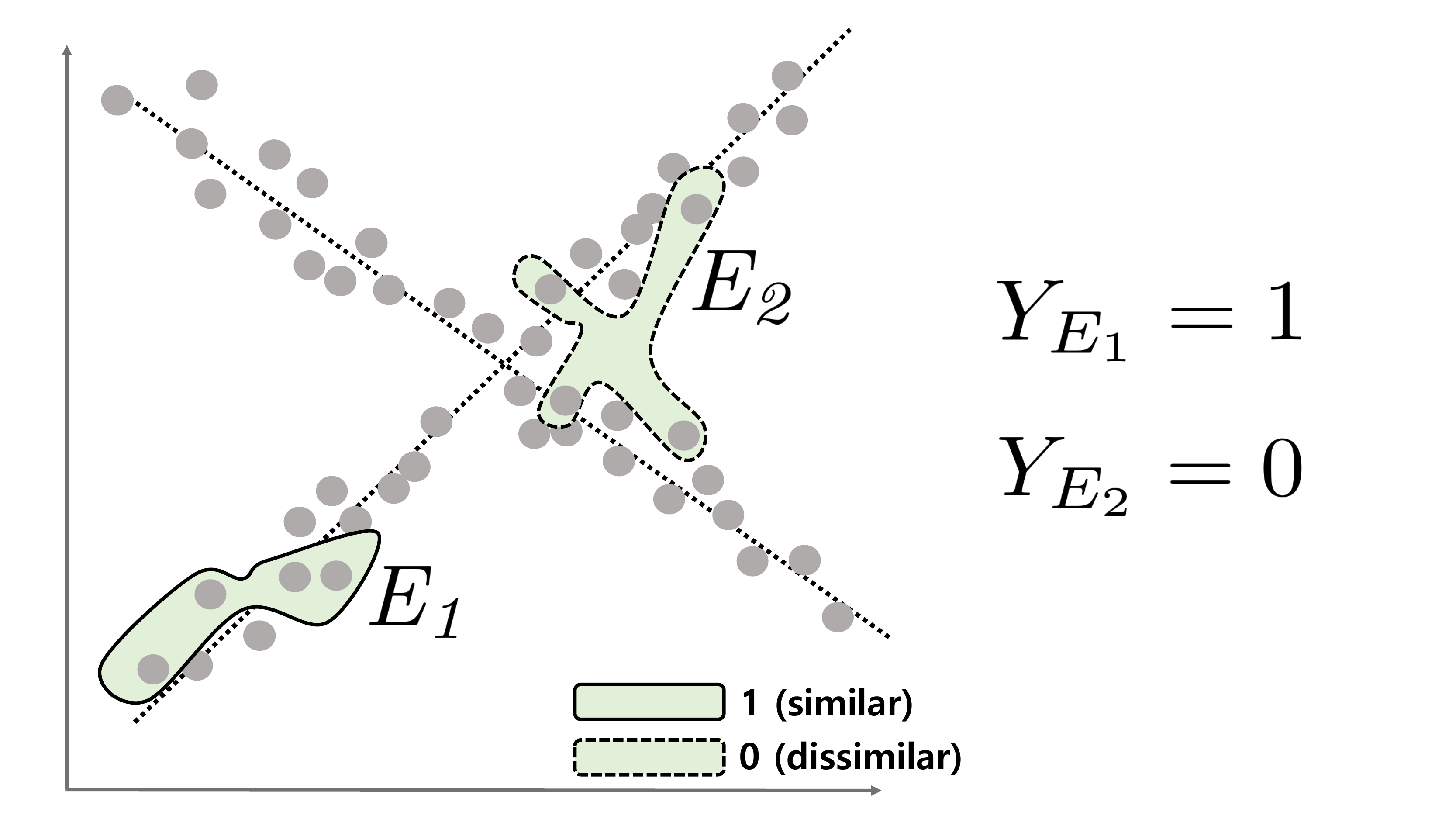}
		\caption{\footnotesize{\textbf{Connection to subspace clustering.} Subspace clustering is illustrated for a simple scenario in which the entire signal space is two-dimensional and data points are approximately lying on a union of two $1$-dimensional affine spaces (lines). A common procedure in the existing algorithms includes construction of a $d$-th order affinity tensor ($d\geq 2$) each entry of which represents a quantity that captures a level of similarity across $d$ data points, so taking either 0 or 1 depending on the similarity level. For instance, the four points involved in $E_1$ in the figure lie near the same affine space, so the similarity measure is decided as $1$; on the other hand, the four points in $E_2$ span different affine spaces, so the similarity measure is decided as $0$.
		Since each data point does not exactly lie in a subspace, an error can occur in the decision---the similarity measurement can be noisy. Hence one can view this problem as the GCBM under the homogeneity measurement model.}
			\label{fig:main1}}
	\end{figure}
	Subspace clustering is a popular problem of which the task is to cluster $n$ data points that approximately lie in a union of lower-dimensional affine spaces.
	The problem arises in a variety of applications such as motion segmentation~\cite{vidal2008multiframe} and face clustering~\cite{ho2003clustering}, where data points corresponding to the same class (tracked points on a moving object or faces of a person) lie on a single lower-dimensional subspace; for details, see~\cite{vidalsurvey} and references therein.
	A common procedure of the existing algorithms for subspace clustering~\cite{chen2009spectral, elhamifar2013sparse, dyer2013greedy,heckel2015robust} begins construction of a $d$-th order affinity tensor ($d\geq 2$) whose entries represent \emph{similarities} between every $d$ data points.
	Since this construction incurs a complexity that scales like $n^d$, sampling-based approaches are proposed in~\cite{govindu2005tensor, chen2009spectral,agarwal2006higher}. 
	
		A similarity between $d$ data points in prior works~\cite{govindu2005tensor, chen2009spectral,agarwal2006higher} is defined such that it tends to $1$ if all of the $d$ points are on the same subspace and $0$ otherwise. Hence, restricted to the two-subspace case, one can view a similarity over a $d$-tuple $E$ as a homogeneity measurement \footnote{In subspace clustering, similarities can be sometimes noisy in that even though the $d$ data points are from the same (different) subspace, similarity can be $0$ ($1$). Note that $Z_E$ in \eqref{def:model} precisely captures this noise.}. 
		By setting the probability of each entry being sampled as $p$, one can relate this to our homogeneity measurement model; see Fig.~\ref{fig:main1} for visual illustration.

 \subsubsection{Channel coding and the parity measurement}
	\begin{figure}
		\centering
		\includegraphics[width=.45\textwidth]{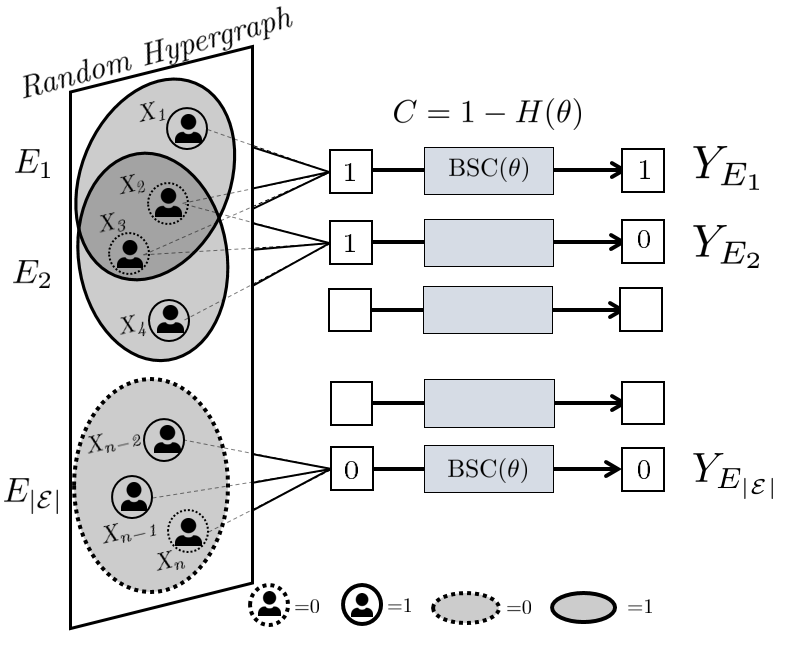}
		\caption{\footnotesize{\textbf{Connection to channel coding.} GCBM with the parity information can be seen as a channel coding problem which employs \rr.
		To see this, we first draw a random $d$-uniform hypergraph with $n$ nodes, where each edge of size $d$ appears with probability $p$. 
		Given the input sequence of $n$ information bits, the parity bits corresponding to all the sampled hyperedges are concatenated, forming a codeword. 
		The noisy measurement can be mapped to the output of a binary symmetric channel (BSC) with crossover probability $\theta$, when fed by the codeword.
		A recovery algorithm $\psi$ corresponds to the decoder which wishes to infer the $n$ information bits from the received signals. 
		One can then see that recovering communities in hypergraphs is equivalent to the above channel coding problem.}
			\label{fig:main}}
	\end{figure}
	
	The community recovery problem has an inherent connection with channel coding problems~\cite{abbe2014decoding, abbe2016exact}. 
To see this, consider a communication setting which employs \rr. 
	To make a connection, we begin by constructing a random $d$-uniform hypergraph with $n$ nodes, where each edge of size $d$ appears with probability $p$. 
	Given the input sequence of $n$ information bits, we then concatenate the parity bits with respect to the sampled hyperedges to form a codeword of average length $p \binom{n}{d}$. 
	Note that the expected code rate is $\frac{n}{p{n \choose d}}$.
The noisy measurement can be mapped to the output of a binary symmetric channel (BSC) with crossover probability $\theta$, when fed by the codeword.
	A recovery algorithm $\psi$ corresponds to the decoder which wishes to infer the $n$ information bits from the received signals. 
	One can then see that recovering communities in hypergraphs is equivalent to the above channel coding problem; see Fig.~\ref{fig:main} for visual illustration.

\section{Main results}
\label{sec:MainResults}
 
\subsection{The homogeneity measurement
	 case}
\begin{theorem}\label{thm:main1}
	Fix $d \geq 2$ and $\epsilon >0$. Under the homogeneity measurement case ($f = f_h$),
	\begin{align*}
		\begin{cases}
			\inf_{\psi}P_e(\psi) \to 0 & \text{if}~ \binom{n}{d}p \geq (1+\epsilon)\frac{2^{d-2}}{d} \frac{n \log n}{(\sqrt{1-\theta}-\sqrt{\theta})^2}; \\
			\inf_{\psi}P_e(\psi) \not\rightarrow 0 &\text{if}~ \binom{n}{d}p \leq (1-\epsilon)\frac{2^{d-2}}{d} \frac{n\log n}{(\sqrt{1-\theta}-\sqrt{\theta})^2}.
		\end{cases}
	\end{align*}
\end{theorem}
\begin{IEEEproof} See Sec.~\ref{pf:thm1}.\end{IEEEproof}

We first make a comparison to the result in~\cite{JMLR:v18:16-100}.
While~\cite{JMLR:v18:16-100} models a fairly general similarity measurement, it considers a more relaxed performance metric, so called almost exact recovery, which allows a vanishing fraction of misclassified nodes; and provides a sufficient condition on sample complexity under the setting~\cite{hajek2016information}. 
On the other hand, we identify the sufficient and necessary condition for \emph{exact} recovery, thereby characterizing the fundamental limit.
Specializing their result to the model of our interest, the sufficient condition in~\cite{JMLR:v18:16-100} reads $\Omega(n \log^2n)$, which comes with an extra $\log n$ factor gap to the optimality.

One interesting observation in Theorem~\ref{thm:main1} is that the sample complexity limit is proportional to $\frac{2^{d-2}}{d}$.
This suggests that the amount of information that one hyperedge reveals on average is approximately $\frac{d}{2^{d-2}}$ bits.
To understand why this is the case, consider a setting in which $\theta=0$ and an hyperedge $E=\{i_1,i_2,\cdots, i_d\}$ is observed.
The case of $Y_E=1$ implies $X_{i_1} = X_{i_2} = \cdots = X_{i_d}$, in which there are only two uncertain cases (all zeros and all ones), i.e., the $d-1$ bits of information are revealed. 
On the other hand, the case of $Y_E=0$ provides much less information as it rules out only two possible cases ($X_{i_1} = X_{i_2} = \cdots = X_{i_d} = 0$ and $X_{i_1} = X_{i_2} = \cdots = X_{i_d} = 1$) out of $2^d$ possible candidates. This amounts to roughly $d\cdot \frac{2}{2^{d}}$ bits.
Since $Y_E=1$ occurs with probability $\frac{1}{2^{d-1}}$, the amount of information that one hyperedge can carry on average should read about $\frac{1}{2^{d-1}}(d-1) + \left(1 - \frac{1}{2^{d-1}}\right)\frac{d}{2^{d-1}} \approx \frac{d}{2^{d-2}}$.

Relying on the connection to subspace clustering elaborated in Sec.~\ref{sec:model_just}, one can make an interesting implication from Theorem~\ref{thm:main1}. The result offers a detailed guideline as to how to choose $d$ for sample-efficient subspace clustering.
In the case where the measurement quality reflected in $\theta$ is irrelevant of the number $d$ of data points involved in a measurement, the limit increases in $d$.
In practical applications, however, $\theta$ may depend on $d$. Actually, the quality of similarity measure can improve as more data points get involved, making $\theta$ decrease as $d$ increases. 
In this case, choosing $d$ as small as possible minimizes $\frac{2^{d-2}}{d}$ but may make $\theta$ too large.
Hence, there might be a \emph{sweet spot} on $d$ that minimizes the sample complexity.
It turns out this is indeed the case in practice. 
Actually we identify such optimal $d^*$ for motion segmentation application; see Sec.~\ref{sec:exhom} for details.

\subsection{The parity measurement case}

\begin{theorem}
	\label{thm:main2}
Fix $d \geq 2$ and $\epsilon >0$. Under the parity measurement case ($f = f_p$),
	\begin{align*}
	\begin{cases}
\inf_{\psi}P_e (\psi) \rightarrow 0 & \text{if}~ \binom{n}{d}p \geq  (1+\epsilon)\frac{1}{d}\frac{n\log n}{(\sqrt{1-\theta}-\sqrt{\theta})^2}\,; \\
	\inf_{\psi}P_e (\psi) \not\rightarrow 0 & \text{if}~ \binom{n}{d}p \leq (1+\epsilon)\frac{1}{d}\frac{n\log n}{(\sqrt{1-\theta}-\sqrt{\theta})^2}\,.
	\end{cases}
	\end{align*}
\end{theorem}
\begin{IEEEproof} See Sec.\ref{pf:thm2}.\end{IEEEproof}

Notice that for a fixed $\theta$ and $n$, the minimum sample complexity is proportional to $\frac{1}{d}$, hence decreases in $d$ unlike \hh~case. 

In view of the connection made in Sec.~\ref{sec:model_just}, a natural question that arises in the context of channel coding is to ask how far the rate of the random LDGM code is from the capacity of the BSC channel.
The connection can help immediately answer the question. 
We see from Theorem~\ref{thm:main2} that the rate of the LDGM code is
\begin{align*}
\frac{n}{p{n \choose d}} = \frac{d(\sqrt{1-\theta}-\sqrt{\theta})^2}{\log n}.
\end{align*} 
This suggests that the code rate increases in $d$. Note that as long as $d$ is constant, the rate vanishes, being far from the capacity of BSC channel $1-H(\theta)$.
On the other hand, it is not clear as to whether or not the random LDGM code can achieve a non-vanishing code rate possibly by increasing the value of $d$.
 To check this, we explore the case where $d$ can scale with $n$. 
By symmetry, it suffices to consider the case $2\leq d\leq n/2$.
Moreover, to avoid pathological cases where $d$ fluctuates as $n$ increases, we assume that $d$ is a monotone function. 
\begin{theorem}
\label{thm:main3}
Fix $d$, a monotone function of $n$ such that $2\leq d \leq n/2$, and $\epsilon > 0$.
Under the parity measurement case ($f = f_p$),
\begin{itemize}
	\item(upper bound) $\inf_{\psi}P_e (\psi) \rightarrow 0$ if 
	\begin{align}
	\binom{n}{d}p &\geq (1+\epsilon) \frac{5/2}{d}\frac{n\log n}{(\sqrt{1-\theta}-\sqrt{\theta})^2}~\text{and} \label{ubb1}\\
	\binom{n}{d}p &\geq (1+\epsilon) 5\log 2\frac{n}{(\sqrt{1-\theta}-\sqrt{\theta})^2}\,;\label{ubb2}
	 \intertext{\item(lower bound)  $\inf_{\psi}P_e (\psi) \not\rightarrow 0$ if}
	\binom{n}{d} p &\leq (1-\epsilon) \frac{1}{d} \frac{n\log n }{(\sqrt{1-\theta}-\sqrt{\theta})^2}~\text{or}\label{lbb1}\\
	\binom{n}{d} p &\leq \frac{n}{1-H(\theta)}\,. \label{lbb2}
	\end{align}
\end{itemize}
\end{theorem}
\begin{IEEEproof} See Sec.~\ref{pf:thm3}.\end{IEEEproof}

To see what these results mean, consider the two cases: $d = \Omega(\log n)$ and $d = o(\log n)$.
In the case $d = \Omega(\log n)$, the theorem says that for a fixed $\theta$,
\begin{align*}
\inf_{\psi}P_e(\psi)\to 0 &\text{ if } \binom{n}{d}p > \beta_1 n~\text{and}\\
\inf_{\psi}P_e(\psi)\not\to 0 &\text{ if } \binom{n}{d}p < \beta_2 n
\,,
\end{align*} where $\beta_1 = \max\left\{\frac{5/2 \log n}{(\sqrt{1-\theta}-\sqrt{\theta})^2 d},~ \frac{5\log 2}{(\sqrt{1-\theta}-\sqrt{\theta})^2} \right\}\asymp 1$ and $\beta_2=\max\left\{ \frac{\log n}{(\sqrt{1-\theta}-\sqrt{\theta})^2d},~ \frac{1}{1-H(\theta)}\right\}\asymp 1$. 
This suggests that as long as $d$ grows asymptotically larger than $\log n$, we can  achieve an order-wise tight sample complexity that is linear in $n$.
 On the other hand, in the case $d = o(\log n)$, the theorem asserts that \begin{align*}
\inf_{\psi}P_e(\psi)\to 0 &\text{ if }\binom{n}{d}p >\frac{5/2}{d} \frac{n\log n}{(\sqrt{1-\theta}-\sqrt{\theta})^2}~\text{and}\\
\inf_{\psi}P_e(\psi) \not\to 0 &\text{ if }\binom{n}{d}p < \frac{1}{d} \frac{n\log n }{(\sqrt{1-\theta}-\sqrt{\theta})^2}\,.
\end{align*} 
This implies that one cannot achieve the linear-order sample complexity if $d$ grows slower than $\log n$. 
 The implication of the above two can be formally stated as follows.

\begin{corollary}
	\label{thm:main_dstar}
	For $d=o(\log n)$, reliable recovery is impossible with linear-order sample complexity, while it is possible for $d=\Omega(\log n)$. 
\end{corollary}

From this, we see that the random LDGM code can achieve a constant rate as soon as $d=\Omega(\log n)$.

\section{Proof of Theorem~\ref{thm:main1}}
\label{pf:thm1}

The achievability and converse proofs are streamlined with the help of Lemmas~\ref{lem:bound} and \ref{lem:ind}, of which the proofs are left in Appendix~\ref{appenA}. 
For illustrative purpose, we focus on the noisy case $(\theta > 0)$ and assume that $n$ is even.
For a vector $\mathbf{V}:=\{V_i\}_{1\leq i\leq n} \in \{0,1\}^{n}$, we define
\begin{align}
\begin{cases}
f_{\{i_1,i_2,\cdots, i_d \}} ({\bf V}) &:=  f(V_{i_1}, V_{i_2},\cdots , V_{i_d}); \\
\Ye(\mathbf{V})& :=  \{\ye(\mathbf{V})\}_{E\in \mathcal{E}};\\
{\sf d_H}(\mathbf{V}) & :=  \|\mathbf{Y}-\Ye\mathbf{(V)}\|_1\,.
\end{cases}\label{defs}
\end{align}
Let $\psi_{\text{ML}}$ be the maximum likelihood (ML) decoder.
 One can easily verify that 
\begin{align*}
\psi_\text{ML}(\mathbf{Y}) = \arg \min _{\mathbf{V} \in \{0,1\}^{n}} {\sf d_H}(\mathbf{V}),
\end{align*}
where ties are randomly broken.

\subsection{Achievability proof}
 We intend to prove that
\begin{align*}
\max_{\mathbf{X}\in \{0,1\}^{n}} \Pr(\psi_{\text{ML}}(\mathbf{Y})\notin  \{\mathbf{X},\mathbf{X}\oplus \mathbf{1}\} ) \rightarrow 0
\end{align*}
under the claimed condition.
Let $ \mathbf{A} \in \{0,1\}^n$ be the ground-truth vector. Without loss of generality, assume that the first $k$ coordinates are $0$'s and the next $n-k$ coordinates are $1$'s, where $ 0 \leq k \leq n/2$.  

Let $\mathcal{A}_{i,j}$ denote the collection of all vectors whose coordinates are different from that of $\mathbf{A}$ in $i$ many positions among the first $k$ coordinates and in $j$ many positions among the next $n-k$ coordinates.
Note that $\mathcal{A}_{0,0} =\{\mathbf{A}\}$ and $\mathcal{A}_{k,n-k} =\{\mathbf{A}\oplus \mathbf{1}\}$.
Thus, a decoding algorithm $\psi$ is successful if and only if the output $\psi(\mathbf{Y}) \in\mathcal{A}_{0,0} \cup \mathcal{A}_{k,n-k}$.
Let $\mathcal{I}:=\{(i,j)~:~ (i,j)\notin \{ (0,0), (k,n-k)\},~0\leq i \leq k,~ \text{and}~0\leq j\leq n-k \} $.
We also define
\begin{align*}
\mathbf{V}_{i,j} := ( \underbrace{\underbrace{1,\cdots,1}_{i},0,\cdots, 0}_{k}, \underbrace{\underbrace{0,\cdots ,0}_{j},1,\cdots,1}_{n-k} )\,,
\end{align*}
which is a representative vector of $\mathcal{A}_{i,j}$.

Using these notations and the union bound, we get:
\begin{align}
	&\Pr(\psi_{\text{ML}}(\mathbf{Y})\notin \{\mathbf{X},\mathbf{X}\oplus \mathbf{1}\}~|~ \mathbf{X} =\mathbf{A} ) \nonumber \\
	&\overset{(a)}{\leq}    \Pr\left(\bigcup_{(i,j)\in \mathcal{I}}\bigcup_{ \mathbf{V} \in \mathcal{A}_{i,j}} \left[ {\sf d_H}(\mathbf{V}) \leq {\sf d_H}(\mathbf{A}) \right] \right) \nonumber\\
	&\leq \sum_{(i,j)\in \mathcal{I}}~~\sum_{\mathbf{V} \in \mathcal{A}_{i,j}}\Pr\left( {\sf d_H}(\mathbf{V}) \leq {\sf d_H}(\mathbf{A})  \right) \nonumber\\
	&= \sum_{(i,j)\in \mathcal{I}} \binom{k}{i} \binom{n-k}{j}\Pr\left( {\sf d_H}(\mathbf{V}_{i,j}) \leq {\sf d_H}(\mathbf{A})  \right) \label{upperbound},
\end{align}
where  the step ($a$) follows from the fact that the ML decoder outputs $\mathbf{V} \notin \{\mathbf{A},\mathbf{A}\oplus \mathbf{1}\}$ if ${\sf d_H}(\mathbf{V}) \leq {\sf d_H}(\mathbf{A})$.

To compare ${\sf d_H}(\mathbf{V}_{i,j})$ with ${\sf d_H}(\mathbf{A})$, we define the set of \emph{distinctive} hyperedges, i.e., the set of hyperedges such that $\ye (\mathbf{A}) \neq \ye (\mathbf{V}_{i,j})$:
 \begin{align}
\mathcal{F}_{i,j}:= \left\{ E\in \binom{[n]}{d} ~:~\ye(\mathbf{A})\neq \ye(\mathbf{V}_{i,j}) \right\} \label{comparison}
\end{align} and $\mathcal{E}_{i,j}: = \mathcal{E} \cap \mathcal{F}_{i,j}$.
By definition, for $E\in\mathcal{E}_{i,j} $, $Y_E =\ye(\mathbf{A})$ if $Z_E =0$; $Y_E =\ye(\mathbf{V}_{i,j})$ otherwise. Hence, ${\sf d_H}(\mathbf{V}_{i,j}) \leq {\sf d_H}(\mathbf{A})$ if and only if $ \sum_{E\in \mathcal{E}_{i,j} } Z_E \geq \frac{|\mathcal{E}_{i,j}|}{2}$. This leads to:
\ifdefined\SingleColumn 
	\begin{align}
		&\Pr\left( {\sf d_H}(\mathbf{V}_{i,j}) \leq {\sf d_H}(\mathbf{A})  \right)\nonumber\\ &= \sum_{\ell=1}^{|\mathcal{F}_{i,j}|}\Pr\left( {\sf d_H}(\mathbf{V}_{i,j}) \leq {\sf d_H}(\mathbf{A}) ~| ~ |\mathcal{E}_{i,j}| = \ell\right) \Pr(|\mathcal{E}_{i,j}| = \ell) \label{expansion}\\
		&=\sum_{\ell=1}^{|\mathcal{F}_{i,j}|}\Pr\left( \sum_{E\in \mathcal{E}_{i,j} } Z_E \geq \frac{\ell}{2} ~\bigg| ~ |\mathcal{E}_{i,j}| = \ell\right)\cdot \binom{|\mathcal{F}_{i,j}|}{\ell}p^{\ell}(1-p)^{|\mathcal{F}_{i,j}|-\ell} \nonumber\\
		&\overset{(a)}{\leq}  \sum_{\ell=1}^{|\mathcal{F}_{i,j}|} e^{-\ell D(0.5\| \theta)} \binom{|\mathcal{F}_{i,j}|}{\ell}p^{\ell}(1-p)^{|\mathcal{F}_{i,j}|-\ell} \nonumber\\
		&= (1-(1-e^{-D(0.5\| \theta )})p)^{|\mathcal{F}_{i,j}|}\,, \label{exp1}	
	\end{align} 
\else 
	\begin{align}
		&\Pr\left( {\sf d_H}(\mathbf{V}_{i,j}) \leq {\sf d_H}(\mathbf{A})  \right)\nonumber\\ &= \sum_{\ell=1}^{|\mathcal{F}_{i,j}|}\Pr\left( {\sf d_H}(\mathbf{V}_{i,j}) \leq {\sf d_H}(\mathbf{A}) ~| ~ |\mathcal{E}_{i,j}| = \ell\right) \Pr(|\mathcal{E}_{i,j}| = \ell) \label{expansion}\\
		&=\sum_{\ell=1}^{|\mathcal{F}_{i,j}|}\Pr\left( \sum_{E\in \mathcal{E}_{i,j} } Z_E \geq \frac{\ell}{2} ~\bigg| ~ |\mathcal{E}_{i,j}| = \ell\right)\nonumber\\
		&~~~~~~~~~~~~~~~~~~~~~~~~~~~~~~~~\cdot \binom{|\mathcal{F}_{i,j}|}{\ell}p^{\ell}(1-p)^{|\mathcal{F}_{i,j}|-\ell} \nonumber\\
		&\overset{(a)}{\leq}  \sum_{\ell=1}^{|\mathcal{F}_{i,j}|} e^{-\ell D(0.5\| \theta)} \binom{|\mathcal{F}_{i,j}|}{\ell}p^{\ell}(1-p)^{|\mathcal{F}_{i,j}|-\ell} \nonumber\\
		&= (1-(1-e^{-D(0.5\| \theta )})p)^{|\mathcal{F}_{i,j}|}\,, \label{exp1}	
	\end{align} 
\fi

where ($a$) is due to Chernoff-Hoeffding~\cite{hoeffding1963probability}. 
By letting $p' := (1-e^{-D(0.5\| \theta )})p$ and applying this to \eqref{upperbound}, we get:
\begin{align}
	&\Pr(\psi_{\text{ML}}(\mathbf{Y})\notin \{\mathbf{X},\mathbf{X}\oplus \mathbf{1}\}~|~ \mathbf{X} =\mathbf{A} ) \nonumber\\
	&\leq \sum_{(i,j)\in \mathcal{I}} \binom{k}{i} \binom{n-k}{j}(1-p')^{|\mathcal{F}_{i,j}|} \label{upperbound2}.
\end{align}  

To give a tight upper bound on \eqref{upperbound2}, one needs a tight lower bound on the size of the set of distinctive hyperedges, i.e., $|\mathcal{F}_{i,j}|$. 
It turns out that bounding $|{\cal F}_{i,j}|$ when $d > 2$ requires non-trivial combinatorial counting.
Note that this was not the case when $d = 2$ since $|{\cal F}_{i,j}|$ can be exactly computed via simple counting.
Indeed, one of our main technical contributions lies in the derivation of tight bounds on $|{\cal F}_{i,j}|$, which we detail below.

\begin{fact} \label{fact1} The number of distinctive hyperedges can be calculated as follows:
\ifdefined\SingleColumn 
	\begin{align}
		&|\mathcal{F}_{i,j}| =  \sum_{\ell=1}^{d-1} \binom{i}{\ell}\binom{k-i}{d-\ell} +\sum_{\ell=1}^{d-1} \binom{j}{\ell}\binom{n-k-j}{d-\ell} +\sum_{\ell=1}^{d-1} \binom{i}{\ell}\binom{n-k-j}{d-\ell}+ \sum_{\ell=1}^{d-1} \binom{k-i}{\ell}\binom{j}{d-\ell}. \label{count}
	\end{align}
\else
	\begin{align}
		&|\mathcal{F}_{i,j}| =  \sum_{\ell=1}^{d-1} \binom{i}{\ell}\binom{k-i}{d-\ell} +\sum_{\ell=1}^{d-1} \binom{j}{\ell}\binom{n-k-j}{d-\ell} \nonumber
		\\&+\sum_{\ell=1}^{d-1} \binom{i}{\ell}\binom{n-k-j}{d-\ell}+ \sum_{\ell=1}^{d-1} \binom{k-i}{\ell}\binom{j}{d-\ell}. \label{count}
	\end{align}
\fi
	
\end{fact}
\begin{IEEEproof}
Consider a hyperedge $E = \{i_1, i_2, \cdots, i_d\}$ such that $f_E(\mathbf{A}) = 1$. That is, the hyperedge is connected only to a subset of the first $k$ nodes or only to a subset of the last $n-k$ nodes. 
That is, $\{i_1, i_2, \cdots, i_d\} \subset \{1,2,\cdots, k\}$ or $\{i_1, i_2, \cdots, i_d\} \subset \{k+1,k+2,\cdots, n\}$.
Consider the first case, i.e., $\{i_1, i_2, \cdots, i_d\} \subset \{1,2,\cdots, k\}$. 
In order for this hyperedge to be distinctive, i.e., $f_E(\mathbf{V}_{i,j}) = 0$, at least one element of $E$ must be in $\{1,2,\cdots,i\}$, and at least one element of $E$ must be in $\{i+1, \cdots, k\}$.
Thus, the total number of such distinctive hyperedges is $\sum_{\ell=1}^{d-1} \binom{i}{\ell}\binom{k-i}{d-\ell}$.
Similarly, one can count the number of distinctive hyperedges for the case $\{i_1, i_2, \cdots, i_d\} \subset \{k+1,k+2,\cdots, n\}$: $\sum_{\ell=1}^{d-1} \binom{j}{\ell}\binom{n-k-j}{d-\ell}$.
By considering the opposite case where $f_E(\mathbf{A}) = 0$ and $f_E(\mathbf{V}_{i,j}) = 1$, one can also obtain the remaining two terms, proving the statement. \end{IEEEproof}

By symmetry, we see that  $|\mathcal{F}_{i,j}| = |\mathcal{F}_{k-i,n-k-j}|$.
Hence,
\begin{align}
&\sum_{(i,j)\in \mathcal{I}} \binom{k}{i} \binom{n-k}{j}(1-p')^{|\mathcal{F}_{i,j}|} \\
&\leq \sum_{(i,j)\in \mathcal{I},~j\leq\lfloor\frac{n-k}{2} \rfloor} \binom{k}{i} \binom{n-k}{j}(1-p')^{|\mathcal{F}_{i,j}|} + \sum_{(i,j)\in \mathcal{I},~j\geq\lceil\frac{n-k}{2} \rceil} \binom{k}{i} \binom{n-k}{j}(1-p')^{|\mathcal{F}_{i,j}|}\\
&= \sum_{(i,j)\in \mathcal{I},~j\leq\lfloor\frac{n-k}{2} \rfloor} \binom{k}{i} \binom{n-k}{j}(1-p')^{|\mathcal{F}_{i,j}|} + \sum_{(i,j)\in \mathcal{I},~j\leq\lfloor\frac{n-k}{2} \rfloor} \binom{k}{k-i} \binom{n-k}{n-k-j}(1-p')^{|\mathcal{F}_{k-i,n-k-j}|}\\
&= 2\sum_{(i,j)\in \mathcal{I},~j\leq\lfloor\frac{n-k}{2} \rfloor} \binom{k}{i} \binom{n-k}{j}(1-p')^{|\mathcal{F}_{i,j}|} =: 2V.
\end{align}

In order to bound $V$, for a fixed constant $\delta >0$, we define the following index sets: $\mathcal{I}_\text{big} := \{(i,j)\in \mathcal{I} : \left[j \leq \frac{n-k}{2}\right] \cap \left(\left[i \geq \delta n\right] \cup \left[j \geq \delta n\right]\right)\}$ and $\mathcal{I}_\text{small} := \{(i,j)\in \mathcal{I} : \left[j \leq \frac{n-k}{2}\right] \cap \left(\left[i < \delta n\right] \cap \left[j < \delta n\right]\right)\}$.
Then, 
\begin{align}
V &= \sum_{(i,j)\in \mathcal{I}_\text{big} \cup \mathcal{I}_\text{big}}  \binom{k}{i} \binom{n-k}{j}(1-p')^{|\mathcal{F}_{i,j}|}\\
&= \sum_{(i,j)\in \mathcal{I}_\text{big}}  \binom{k}{i} \binom{n-k}{j}(1-p')^{|\mathcal{F}_{i,j}|} \label{upperfirst}\\
&+ \sum_{(i,j)\in \mathcal{I}_\text{small}}  \binom{k}{i} \binom{n-k}{j}(1-p')^{|\mathcal{F}_{i,j}|}.\label{uppersec}
\end{align}
Let us first consider \eqref{upperfirst}. 
Without loss of generality, assume $i\geq \delta n$.
Then it follows from \eqref{count} that 
\begin{align*}
	|\mathcal{F}_{i,j}| &\geq \sum_{\ell=1}^{d-1} \binom{i}{\ell}\binom{n-k-j}{d-\ell} \overset{(a)}{\geq} \sum_{\ell=1}^{d-1} \binom{i}{\ell}\binom{n/4}{d-\ell}\\ 
	&\geq\binom{i}{1}\binom{n/4}{d-1} \geq \delta n \binom{n/4}{d-1} = \Omega(n^d), 
\end{align*}
where ($a$) follows from the hypothesis that $j\leq \frac{n-k}{2}$ and $k\leq \frac{n}{2}$.
Then it is easy to show that \eqref{upperfirst}$\to 0$:
\begin{align*}
	\eqref{upperfirst} &\leq \sum_{(i,j)\in \mathcal{I}} \binom{k}{i} \binom{n-k}{j}e^{-p'\Omega(n^d)} \\
	&\overset{(a)}{=}e ^{-\Omega(n\log n )}\sum_{(i,j)\in \mathcal{I}} \binom{k}{i} \binom{n-k}{j}\leq e ^{-\Omega(n\log n )} 2^{n} \to 0, 
\end{align*}
where ($a$) follows from the fact that $p'\Omega(n^d)\asymp p\binom{n}{d} =\Omega(n\log n)$.

Now we consider \eqref{uppersec}. 
The following lemma gives a tight lower bound on $|\mathcal{F}_{i,j}|$ for this case:
\begin{lemma}\label{lem:bound}
	For $i< \delta n$ and $j<\delta n $,
	\[
	|\mathcal{F}_{i,j}|\geq (i+j)\cdot\frac{(1-2\delta)^{d-1}}{2^{d-2}}  \binom{n-1}{d-1}.
	\]
\end{lemma}
\begin{IEEEproof}
	See Sec.~\ref{app1}.
	\end{IEEEproof}
Applying Lemma~\ref{lem:bound} to \eqref{uppersec}, we get:
\begin{align}
	\eqref{uppersec}
	&=\sum_{(i,j)\in \mathcal{I}_\text{small}}\binom{k}{i} \binom{n-k}{j}(1-p')^{|\mathcal{F}_{i,j}|}\nonumber\\
	&\overset{(a)}{\leq} \sum_{(i,j)\in \mathcal{I}_\text{small}} n^i n^j e^{-p'(i+j)\cdot\frac{(1-2\delta)^{d-1}}{2^{d-2}}  \binom{n-1}{d-1}} \nonumber\\
	&= \sum_{(i,j)\in \mathcal{I}_\text{small}} \exp\left((i+j)\left\{\log n-\frac{p'(1-2\delta)^{d-1}\binom{n-1}{d-1}}{2^{d-2}} \right\} \right), \label{c1}
\end{align}
where ($a$) follows due to $\binom{k}{i}\leq n^i$, $\binom{n-k}{j}\leq n^j$ and Lemma~\ref{lem:bound}.
A straightforward computation yields $(1-e^{-{\sf D_{KL}}(0.5\| \theta )})= (\sqrt{1-\theta}-\sqrt{\theta})^2$, so the claimed condition \begin{align*}
\binom{n}{d}p \geq (1+\epsilon)\frac{2^{d-2}}{d} \frac{n \log n}{(\sqrt{1-\theta}-\sqrt{\theta})^2}
\end{align*} becomes 
\begin{align}
	\binom{n}{d}p' \geq (1+\epsilon) \frac{2^{d-2}}{d} n\log n\,.\label{suff}
\end{align}
Under the claimed condition, we get:
\begin{align*}
	\frac{p'(1-2\delta)^{d-1}\binom{n-1}{d-1}}{2^{d-2}} &=\frac{p'(1-2\delta)^{d-1}\binom{n}{d}\frac{d}{n}}{2^{d-2}}\\
	&\overset{(a)}{\geq} (1+\epsilon)(1-2\delta)^{d-1} \log n\\
	&\overset{(b)}{\geq} (1+\epsilon/2) \log n,
\end{align*}
where ($a$) follows from \eqref{suff}; ($b$) follows by choosing $\delta $ sufficiently small ($(1-2\delta)^{d-1}\to 0$ as $\delta\to 0$).
Thus, \eqref{c1} converges to $0$ as $n$ tends to infinity. This completes the proof.

\subsection{Converse proof}
Let $\mathcal{V}_{1/2}$ be the collection of $n$-dimensional vectors, each consisting of $n/2$ number of $0$'s and $n/2$ number of $1$'s.   Moreover, let $\mathbf{X}_{1/2}$ be the random vector sampled uniformly at random over $\mathcal{V}_{1/2}$. For any scheme $\psi$, by definition of $P_e (\psi)$, we see that 
\begin{align*}
	\Pr\left(\psi(\mathbf{Y}) \notin \{\mathbf{X},~ \mathbf{X}\oplus \mathbf{1}\}~|~ \mathbf{X} = \mathbf{X}_{1/2} \right)  \leq P_e(\psi)
\end{align*}
and hence
\begin{align*}
	\inf_{\psi } \Pr\left(\psi(\mathbf{Y}) \notin \{\mathbf{X},~ \mathbf{X}\oplus \mathbf{1}\}~|~ \mathbf{X} = \mathbf{X}_{1/2} \right)  \leq \inf_{\psi }P_e(\psi).
\end{align*}
Relying on this inequality, our proof strategy is to show that the left hand side is strictly bounded away from $0$.
Note that the infimum in the left hand side is achieved by $\psi_{\text{ML},1/2}$:
\begin{align*}	\psi_{\text{ML},1/2}(\mathbf{Y}) = \arg \min _{\mathbf{V} \in \mathcal{V}_{1/2}} {\sf d_H}(\mathbf{V})\,.
\end{align*}	
By letting $\mathbf{A} = ( \underbrace{0,\cdots, 0}_{n/2}, \underbrace{1,\cdots,1}_{n/2} )$, 
we obtain 
\begin{align*}
	&\Pr\left(\psi_{\text{ML},1/2}(\mathbf{Y}) \notin \{\mathbf{X},~ \mathbf{X}\oplus \mathbf{1}\}~|~ \mathbf{X} = \mathbf{X}_{1/2} \right)\\
	= &\Pr\left(\psi_{\text{ML},1/2}(\mathbf{Y}) \notin \{\mathbf{A},~ \mathbf{A}\oplus \mathbf{1}\}~|~ \mathbf{X} = \mathbf{A}\right).
\end{align*}
Let $S$ be the success event:
\begin{align*}
S :=\bigcap_{\mathbf{V}\in \mathcal{V}_{1/2}\setminus \{\mathbf{A}, \mathbf{A}\oplus \mathbf{1}\} } \left[{\sf d_H}(\mathbf{V})> {\sf d_H}(\mathbf{A})\right]\,.
\end{align*}
One can show that $\Pr\left(\psi_{\text{ML},1/2}(\mathbf{Y}) \notin \{\mathbf{A},~ \mathbf{A}\oplus \mathbf{1}\}~|~ \mathbf{X} = \mathbf{A}\right) \geq \frac{1}{3} \Pr(S^c)$.
This is due to the fact that given $S^c$, there are more than two candidates for $\arg\min_{\mathbf{V}\in \mathcal{V}_{1/2}}{\sf d_H}(\mathbf{V})$, so 
\[
\Pr\left(\psi_{\text{ML},1/2}(\mathbf{Y}) \notin \{\mathbf{A},~ \mathbf{A}\oplus \mathbf{1}\}~|~ \mathbf{X} = \mathbf{A},~S^c\right) \geq \frac{1}{3}.
\]
Hence, it suffices to show $\Pr(S)\to 0$.
To give a tight upper bound on $\Pr(S)$, we construct a subset of nodes  such that any two nodes in the subset do not share the same hyperedge. To this end, we use the deletion technique (alteration technique)~\cite{alon2004probabilistic}.
We first choose a big subset 
\[
\mathcal{R}_{\text{big}}=\left\{1,2,\cdots,r \right\} \bigcup \left\{\frac{n}{2}+1,\frac{n}{2}+2,\cdots, \frac{n}{2}+r  \right\},
\]
where $r=\lceil \frac{n}{\log^7 n} \rceil$; then erase every node in $\mathcal{R}_{\text{big}}$ which shares hyperedges with other nodes in $\mathcal{R}_{\text{big}}$ to obtain $\mathcal{R}_{\text{res}}$. The following lemma guarantees that $\mathcal{R}_{\text{res}}$ has a comparable size as that of $\mathcal{R}_{\text{big}}$ with high probability. For the later usage, we allow $d$ to scale with $n$. 
\begin{lemma} \label{lem:ind}
	Suppose $\binom{n}{d}p=O(n\log n)$ and $d=O(\log n)$. Let $\mathcal{R}_{\text{big}}$ be a subset of $[n]$ and $\mathcal{R}_{\text{res}}$ be a subset obtained from $\mathcal{R}_{\text{big}}$ by deleting every node which shares hyperedges with other nodes in $\mathcal{R}_{\text{big}}$.
	If $|\mathcal{R}_{\text{big}}| = O(n/\log^7 n)$, then 
	with probability approaching $1$, 
	\begin{align*} 
		|\mathcal{R}_{\text{res}}|=(1-o(1))|\mathcal{R}_{\text{big}}|\,.
	\end{align*}
\end{lemma}
\begin{IEEEproof}
	See Sec.~\ref{pf:ind}.
\end{IEEEproof}
Let $\Delta$ be the event that $|\mathcal{R}_{\text{res}}|\geq (1-o(1))|\mathcal{R}_{\text{big}}|$.
Given the event $\Delta$, both $\{1,2,\cdots,n/2\} \cap \mathcal{R}_{\text{res}} \label{set1}$ and $\left\{\frac{n}{2}+1,\frac{n}{2}+2,\cdots,n\right\} \cap \mathcal{R}_{\text{res}} \label{set2}$ contain more than $r/2$ elements.
We collect $r/2$ elements from each of these sets and denote by $\{ b_1, b_2 , \cdots, b_{r/2} \}$ and $\{ c_1, c_2 , \cdots, c_{r/2} \}$, respectively. 
Suppose that there exist ($k,\ell$) such that ${\sf d_H}(\mathbf{A} \oplus \mathbf{e}_{b_k}) \leq {\sf d_H}(\mathbf{A})$ and ${\sf d_H}(\mathbf{A} \oplus \mathbf{e}_{c_\ell}) \leq {\sf d_H}(\mathbf{A})$. 
Conditioning on $\Delta$, there are no hyperedges that contain both $b_k$ and $c_\ell$, so ${\sf d_H}(\mathbf{A} \oplus\mathbf{e}_{b_k} \oplus \mathbf{e}_{c_\ell} )\leq {\sf d_H}(\mathbf{A})$. Hence conditioning on $\Delta$,
\begin{align*}
	S &\subset   \bigcap_{k=1}^{r/2} \left[ {\sf d_H}(\mathbf{A} \oplus \mathbf{e}_{b_k}) >{\sf d_H}(\mathbf{A}) \right] \bigcup \bigcap_{k=1}^{r/2} \left[ {\sf d_H}(\mathbf{A} \oplus \mathbf{e}_{c_k}) >{\sf d_H}(\mathbf{A}) \right]\\
	&=:S'.
\end{align*}
Since the event $\Delta$ occurs with probability approaching $1$ and $S \subset S'$, $\Pr(S) \simeq \Pr(S~|~\Delta) \leq \Pr(S'~|~\Delta)$. 
Hence, 
\begin{align*}
	\Pr(S) &\lesssim \Pr\left (S' ~|~ \Delta \right)\\
	&\leq 2\Pr\left ( \bigcap_{k=1}^{r/2} \left[ {\sf d_H}(\mathbf{A} \oplus \mathbf{e}_{b_k}) >{\sf d_H}(\mathbf{A}) \right] ~\bigg|~ \Delta \right) \\ 
	&\overset{(a)}{=} 2\Pr\left (  {\sf d_H}(\mathbf{A} \oplus \mathbf{e}_{b_1}) >{\sf d_H}(\mathbf{A})  ~\big|~ \Delta \right)^{r/2}, 
\end{align*}
where $(a)$ follows from the fact that the events $\{[{\sf d_H}(\mathbf{A} \oplus \mathbf{e}_{b_k}) >{\sf d_H}(\mathbf{A})]\}_{1\leq k \leq r/2 }$ are mutually independent conditioned on $\Delta$.
Let $p' = (1-e^{-{\sf D_{KL}}(0.5\| \theta )})p$ as in the achievability proof. We intend to give an upper bound on $\Pr\left (  {\sf d_H}(\mathbf{A} \oplus \mathbf{e}_{b_1}) >{\sf d_H}(\mathbf{A})  ~\big|~ \Delta \right)$,
i.e., a lower bound on $\Pr\left (  {\sf d_H}(\mathbf{A} \oplus \mathbf{e}_{b_1}) \leq {\sf d_H}(\mathbf{A})  ~\big|~ \Delta \right)$.
Recall from the proof of achievability (see \eqref{exp1}) that
\begin{align*}
 \Pr\left( {\sf d_H}(\mathbf{V}_{i,j}) \leq {\sf d_H}(\mathbf{A})  \right) \leq (1-(1-e^{-{\sf D_{KL}}(0.5\| \theta )})p)^{|\mathcal{F}_{i,j}|}\,.
\end{align*}
For the case of $\mathbf{V}_{i,j} =\mathbf{A} \oplus \mathbf{e}_{b_1}$, $|\mathcal{F}_{i,j}|=  \binom{n/2-1}{d-1} + \binom{n/2}{d-1}$ (note that $k=n/2, i=1, j=0$). So we get:
\begin{align}
\Pr\left( {\sf d_H}(\mathbf{A} \oplus \mathbf{e}_{b_1}) \leq {\sf d_H}(\mathbf{A})  \right) \leq e^{-p'\left( \binom{n/2-1}{d-1} + \binom{n/2}{d-1}\right)}\,. \label{exp:6}
\end{align} 
 On the other hand, what we need for the converse proof is a lower bound. In what follows, we will show that \eqref{exp:6} is tight enough, more precisely, 
\begin{align}
&\Pr\left (  {\sf d_H}(\mathbf{A} \oplus \mathbf{e}_{b_1}) \leq {\sf d_H}(\mathbf{A})  ~\big|~ \Delta \right)\geq (1-o(1))e^{-2p' \binom{n/2-1}{d-1}}\,. \label{tight}
\end{align}
What this means at a high level is that Chernoff-Hoeffding is tight enough.
Let us condition on the event $\Delta$ for the time being. As in \eqref{comparison}, we define the following sets:
\begin{align*}
&\mathcal{F}_{b_1}:= \left\{ E\in \binom{[n]}{d} ~:~ \ye(\mathbf{A})\neq \ye(\mathbf{A} \oplus \mathbf{e}_{b_1})\right\}
\end{align*}
 and $\mathcal{E}_{b_1}: = \mathcal{E} \cap \mathcal{F}_{b_1}$. 
 By definition, for $E\in\mathcal{E}_{b_1}$, $Y_E =\ye(\mathbf{A})$ if $Z_E =0$; $Y_E =\ye(\mathbf{A} \oplus \mathbf{e}_{b_1})$ otherwise. We see that
 \[
 {\sf d_H}(\mathbf{A} \oplus \mathbf{e}_{b_1}) \leq {\sf d_H}(\mathbf{A}) \Leftrightarrow  \sum_{E\in \mathcal{E}_{b_1}} Z_E \geq \frac{|\mathcal{E}_{b_1}|}{2}\,.
 \] 
 Now we want to manipulate $\Pr\left (  {\sf d_H}(\mathbf{A} \oplus \mathbf{e}_{b_1}) \leq {\sf d_H}(\mathbf{A})  ~\big|~ \Delta \right)$ as we did in \eqref{expansion}.
 However, here we need to give a careful attention to the range of summation as $\mathcal{E}_{b_1}$ cannot be equal to $\mathcal{F}_{b_1}$ due to the following reason. Since we conditioned on $\Delta$, no hyperedge in $\mathcal{E}_{b_1}$ intersects $\mathcal{R}_{\text{big}}$ at more than one node (indeed, $b_1$ is the only node where they intersect); in other words, $\mathcal{E}_{b_1}$ is always contained in a proper subset of $\mathcal{F}_{b_1}$: 
\begin{align}
\mathcal{E}_{b_1} &\subset \mathcal{F}_{b_1} \setminus \left\{E\in \binom{[n]}{d}~:~ |E\cap \mathcal{R}_{\text{big}}| \geq 2  \right\} =:\mathcal{G}_{b_1}. \label{def:G}
\end{align}

Now a manipulation similar to~\eqref{expansion} yields:
\begin{align*}
&\Pr\left( {\sf d_H}(\mathbf{A} \oplus \mathbf{e}_{b_1}) \leq {\sf d_H}(\mathbf{A}) ~|~ \Delta\right)\nonumber\\ 
&= \sum_{\ell=1}^{|\mathcal{G}_{b_1}|}\Pr\left({\sf d_H}(\mathbf{A} \oplus \mathbf{e}_{b_1}) \leq {\sf d_H}(\mathbf{A}) ~\big| ~ |\mathcal{E}_{b_1}| = \ell,~\Delta\right) \Pr(|\mathcal{E}_{b_1}| = \ell| \Delta).
\end{align*}
Since the event $\Delta$ is related to the occurrence of edges in 
\begin{align*}
\left\{E\in \binom{[n]}{d}~:~ |E\cap \mathcal{R}_{\text{big}}| \geq 2  \right\}
\end{align*}
and ${\cal E}_{b_1}$ is subject to \eqref{def:G}, 
$\Delta$ and $[|\mathcal{E}_{b_1}|=\ell]$ are independent. 
 Thus, we get:
\begin{align}
&\Pr\left( {\sf d_H}(\mathbf{A} \oplus \mathbf{e}_{b_1}) \leq {\sf d_H}(\mathbf{A}) ~|~ \Delta\right)\nonumber\\ 
&= \sum_{\ell=1}^{|\mathcal{G}_{b_1}|}\Pr\left({\sf d_H}(\mathbf{A} \oplus \mathbf{e}_{b_1}) \leq {\sf d_H}(\mathbf{A}) ~\big| ~ |\mathcal{E}_{b_1}| = \ell, ~\Delta\right) \Pr(|\mathcal{E}_{b_1}| = \ell) \nonumber\\
&=\sum_{\ell=1}^{|\mathcal{G}_{b_1}|}\Pr\left( \sum_{E\in \mathcal{E}_{b_1} } Z_E \geq \frac{\ell}{2} \bigg|  |\mathcal{E}_{b_1}| = \ell\right) \binom{|\mathcal{G}_{b_1}|}{\ell}\frac{p^{\ell}}{(1-p)^{\ell-|\mathcal{G}_{b_1}|}}. \label{e1}	
\end{align} 
By the reverse Chernoff-Hoeffding bound~\cite{hoeffding1963probability}, for a fixed $\delta>0$, there exists $n_{\delta}>0$ such that
\begin{align*}
\Pr\left( \sum_{E\in \mathcal{E}_{b_1} } Z_E \geq \frac{\ell}{2} \bigg|  |\mathcal{E}_{b_1}| = \ell\right) \geq e^{-(1+\delta)\ell {\sf D_{KL}}(0.5\|\theta)}
\end{align*}
for all $\ell \geq n_{\delta}$. Let $g_n$ be a sequence (to be determined) such that $g_n\to \infty$ as $n\to \infty$. For sufficiently large $n$, 
\begin{align}
\eqref{e1}&\geq \sum_{\ell=1}^{|\mathcal{G}_{b_1}|} \binom{|\mathcal{G}_{b_1}|}{\ell}\frac{(e^{-(1+\delta) {\sf D_{KL}}(0.5\|\theta)}p)^{\ell}}{(1-p)^{\ell-|\mathcal{G}_{b_1}|}}  \label{firstex1} \\
&-\sum_{\ell=1}^{g_n-1} \binom{|\mathcal{G}_{b_1}|}{\ell}\frac{(e^{-(1+\delta) {\sf D_{KL}}(0.5\|\theta)}p)^{\ell}}{(1-p)^{\ell-|\mathcal{G}_{b_1}|}}\,.\label{secondex1} 
\end{align} 

Actually one can choose $g_n$ so that \eqref{secondex1} is negligible compared to \eqref{firstex1}. To see this, we consider:
\begin{align}
\frac{\eqref{secondex1}}{ \eqref{firstex1}}  &\leq 
\frac{(1-p)^{|\mathcal{G}_{b_1}|}\sum_{\ell=1}^{g_n-1}\left(
	|\mathcal{G}_{b_1}|\frac{pe^{-(1+\delta){\sf D_{KL}}(0.5\|\theta)}}{1-p}\right)^\ell}{ (1-p)^{|\mathcal{G}_{b_1}|}\sum_{\ell=1}^{|\mathcal{G}_{b_1}|}\binom{|\mathcal{G}_{b_1}|}{\ell} \left(\frac{pe^{-(1+\delta){\sf D_{KL}}(0.5\|\theta)}}{1-p}\right)^\ell} \nonumber\\
&= \frac{\sum_{\ell=1}^{g_n-1}\left(
	|\mathcal{G}_{b_1}|\frac{pe^{-(1+\delta){\sf D_{KL}}(0.5\|\theta)}}{1-p}\right)^\ell}{\left( 1+ \frac{pe^{-(1+\delta){\sf D_{KL}}(0.5\|\theta)}}{1-p}\right)^{|\mathcal{G}_{b_1}|} }\nonumber	\\
&\overset{(a)}{=}   \frac{\sum_{\ell=1}^{g_n-1}\left(
	|\mathcal{G}_{b_1}|\frac{pe^{-(1+\delta){\sf D_{KL}}(0.5\|\theta)}}{1-p}\right)^\ell}{(1+o(1))\exp\left( |\mathcal{G}_{b_1}| \frac{pe^{-(1+\delta){\sf D_{KL}}(0.5\|\theta)}}{1-p}\right) } \nonumber \\
&=: \frac{\sum_{\ell=1}^{g_n-1} q^\ell }{(1+o(1))e^q} \label{thirdex1},
\end{align}
where ($a$) follows from the fact that $\lim_{x\to 0+}\frac{1+x}{e^{x}}=1$, and the last equation is due to the following definition: $q:= |\mathcal{G}_{b_1}| \frac{pe^{-(1+\delta){\sf D_{KL}}(0.5\|\theta)}}{1-p}$.
One can easily verify that
$|\mathcal{F}_{b_1}| =  \binom{n/2-1}{d-1} + \binom{n/2}{d-1}$ and $|\mathcal{G}_{b_1}| =  \binom{n/2-1-r}{d-1} + \binom{n/2-r}{d-1}$.
Since $r = o(n)$, $\lim_{n\to \infty}|\mathcal{G}_{b_1}|/|\mathcal{F}_{b_1}| \to 1$. 
Thus, 
\begin{align}
q &=|\mathcal{G}_{b_1}| \frac{pe^{-(1+\delta){\sf D_{KL}}(0.5\|\theta)}}{1-p}\\
&\asymp |\mathcal{F}_{b_1}| \frac{pe^{-(1+\delta){\sf D_{KL}}(0.5\|\theta)}}{1-p}\asymp n^{d-1} p = \Omega(\log n)\,. \label{div}
\end{align} 

Therefore, if one chooses $g_n=\left\lfloor \log q\right \rfloor$, 
	$$\frac{\eqref{secondex1}}{ \eqref{firstex1}} = \frac{\sum_{\ell=1}^{g_n-1} q^\ell}{e^q} \leq \frac{g_n q^{g_n}}{e^q} \leq \frac{\log q \cdot q^{\log q}}{e^q} = \frac{\log q \cdot e^{(\log q)^2}}{e^q} \rightarrow 0,$$
	and thus $\eqref{secondex1} = o(1)\cdot \eqref{firstex1}$.

Hence, we get:
\begin{align*}
\eqref{e1}&=\eqref{firstex1}-\eqref{secondex1}\\  &\geq (1-o(1))  \sum_{\ell=1}^{|\mathcal{G}_{b_1}|} \binom{|\mathcal{G}_{b_1}|}{\ell}\frac{(e^{-(1+\delta) {\sf D_{KL}}(0.5\|\theta)}p)^{\ell}}{(1-p)^{\ell-|\mathcal{G}_{b_1}|}}\\
& = (1-o(1)) \left(1- (1-e^{-(1+\delta) {\sf D_{KL}}(0.5\|\theta)})p \right)^{|\mathcal{G}_{b_1}|} \\
&\overset{(a)}{\geq} (1-o(1)) \left(1- (1-e^{-(1+\delta){\sf D_{KL}}(0.5\|\theta)})p \right)^{2\binom{n/2}{d-1}} \\
&\overset{(b)}{=} (1-o(1)) \exp\left(- 2\binom{n/2}{d-1} (1-e^{-(1+\delta) {\sf D_{KL}}(0.5\|\theta)})p \right),
\end{align*}	
where ($a$) follows since $|\mathcal{G}_{b_1}|\leq |\mathcal{F}_{b_1}| \leq  2\binom{n/2}{d-1}$; ($b$) follows from the fact that $\lim_{x\to 0+}\frac{1+x}{e^{x}}=1$. As $\delta>0$ can be chosen arbitrarily small, the term $e^{-(1+\delta) {\sf D_{KL}}(0.5\|\theta)}$ can be made arbitrarily close to $e^{- {\sf D_{KL}}(0.5\|\theta)}$, which in turn ensures that the last term is essentially equal to 
\[
(1-o(1)) e^{-2p'\binom{n/2}{d-1}}.
\]
Applying this to the previous upper bound on $\Pr(S)$, we get:
\begin{align*}
	\Pr(S) &\leq \Pr\left (  {\sf d_H}(\mathbf{A} \oplus \mathbf{e}_{b_1}) >{\sf d_H}(\mathbf{A})  ~\big|~ \Delta \right)^{r/2}\nonumber\\
	&\leq \left(1-(1-o(1))e^{-2p' \binom{n/2}{d-1}}\right)^{r/2}\nonumber\\
	&\leq  \exp\left(-(1-o(1))\frac{r}{2} e^{-2p' \binom{n/2}{d-1}} \right)\nonumber\\
	&= \exp\left(-(1-o(1))\frac{n}{2\log ^7 n} e^{-(1+o(1))\cdot \frac{p' d\binom{n}{d}}{2^{d-2}n} } \right),
\end{align*} 
where the last equality follows from the fact that
\begin{align*}
\lim_{n\to \infty}\frac{2p' \binom{n/2}{d-1}}{
	p' d\binom{n}{d}/2^{d-2}n} \to 1~\text{and}~ r=\left\lceil \frac{n}{\log^7 n} \right\rceil.
\end{align*}
The last term converges to $0$ as $p'\leq (1-\epsilon)\frac{2^{d-2}}{d} \frac{n\log n}{\binom{n}{d}}$.

\section{Proof of Theorem~\ref{thm:main2} } \label{pf:thm2}

In this section, we prove a similar statement for the parity measurement case.

\subsection{Achievability proof}
Note that the parity measurement is \emph{symmetric} in a sense that for any two vector $\mathbf{A}$ and $\mathbf{B}$, $\Pr\left(\psi_{\text{ML}}(\mathbf{Y}) \notin \{\mathbf{X},~ \mathbf{X}\oplus \mathbf{1}\}~|~ \mathbf{X} = \mathbf{A}\right) = \Pr\left(\psi_{\text{ML}}(\mathbf{Y}) \notin \{\mathbf{X},~ \mathbf{X}\oplus \mathbf{1}\}~|~ \mathbf{X} = \mathbf{B}\right)$.
Hence, we will prove that
\begin{align*}
\Pr\left(\psi_{\text{ML}}(\mathbf{Y}) \notin \{\mathbf{X},~ \mathbf{X}\oplus \mathbf{1}\}~|~ \mathbf{X} = \mathbf{0}\right) \rightarrow 0
\end{align*}
under the claimed condition.
Conditioning on $\mathbf{X}=\mathbf{0}$,
\begin{align}
&\Pr\left(\psi_{\text{ML}}(\mathbf{Y}) \notin \{\mathbf{0}, \mathbf{1}\}\right) \nonumber\\
&\leq \Pr\left (\bigcup_{\mathbf{A}\neq \mathbf{0},\mathbf{1}}\left[{\sf d_H}(\mathbf{A})\leq {\sf d_H}(\mathbf{0})\right]\right) \nonumber \\
&=\Pr\left(\bigcup_{k=1}^{n-1}\bigcup_{\|\mathbf{A}\|_1 =k}\left[{\sf d_H}(\mathbf{A})\leq  {\sf d_H}(\mathbf{0})\right]\right)  \nonumber\\
&\leq \sum_{k=1}^{n-1} \sum_{\|\mathbf{A}\|_1=k}\Pr\left({\sf d_H}(\mathbf{A})\leq  {\sf d_H}(\mathbf{0})\right)  \nonumber\\
&\overset{(a)}{=} 2\cdot \sum_{k=1}^{n/2} \sum_{\|\mathbf{A}\|_1=k}\Pr\left({\sf d_H}(\mathbf{A})\leq  {\sf d_H}(\mathbf{0})\right)  \nonumber\\
&\overset{(b)}{=}2\cdot\sum^{n/2}_{k=1}\binom{n}{k}\Pr\left({\sf d_H}\left(\sum_{i=1}^k\mathbf{e}_i\right)\leq  {\sf d_H}(\mathbf{0}) \right), \label{eq:3}
\end{align}
where ($a$) follows form the fact that $\Pr\left({\sf d_H}(\mathbf{A})\leq  {\sf d_H}(\mathbf{0})\right) = \Pr\left({\sf d_H}(\mathbf{A}\oplus \mathbf{1})\leq  {\sf d_H}(\mathbf{0})\right) $; ($b$) follows due to symmetry.
To compare ${\sf d_H}\left(\sum_{i=1}^k\mathbf{e}_i\right)$ and ${\sf d_H}(\mathbf{0})$, we define \begin{align*}
\mathcal{F}_{k}:= \left\{ E\in \binom{[n]}{d} ~:~\ye(\mathbf{0})\neq \ye\left(\sum_{i=1}^k\mathbf{e}_i\right) \right\}
\end{align*} and $\mathcal{E}_{k}: = \mathcal{E} \cap \mathcal{F}_{k}$.
As in \eqref{exp1}, we obtain
\begin{align*}
\Pr\left( {\sf d_H}\left(\sum_{i=1}^k\mathbf{e}_i\right) \leq {\sf d_H}(\mathbf{0})  \right)
&\leq (1-(1-e^{-{\sf D_{KL}}(0.5\| \theta )})p)^{|\mathcal{F}_{k}|}\\
&=(1-p')^{|\mathcal{F}_{k}|}	\,,
\end{align*} 
yielding
\begin{align}
\frac{1}{2}\cdot \eqref{eq:3} \leq \sum_{k=1}^{n/2} \binom{n}{k}(1-p')^{|\mathcal{F}_{k}|} \label{ub2}.
\end{align}  
We again count $|\mathcal{F}_{k}|$ in an effort to obtain a tight upper bound on \eqref{ub2}. Notice that $E\in\mathcal{F}_k$ if $|E\cap [k]|$ is odd, and hence
\begin{align}
|\mathcal{F}_k|=\sum_{\substack{i \leq d \\ i\text{ is odd}}}\binom{k}{i}\cdot\binom{n-k}{d-i}\,. \label{counting}
\end{align} 
Let $\delta >0$ be a small constant that will be determined later. For the case $k\geq \delta n$, it follows that 
\begin{align*}
&|\mathcal{F}_{k}| \geq \binom{k}{1} \binom{n-k}{d-1}\geq \delta n \binom{n/2}{d-1} = \Omega(n^d)\,. 
\end{align*}
Then it is easy to show \eqref{ub2}$\to 0$ for this case:
\begin{align*}
& \sum_{k=\delta n}^{n/2} \binom{n}{k}(1-p')^{|\mathcal{F}_{k}|} \leq \sum_{k=\delta n}^{n/2} \binom{n}{k}e^{-p'\Omega(n^d)}\\
&\overset{(a)}{=}e ^{-\Omega(n\log n )}\sum_{k=\delta n}^{n/2} \binom{n}{k} \leq e ^{-\Omega(n\log n )} 2^{n} 	\to 0\,,
\end{align*}
where ($a$) follows from the fact that $p'\Omega(n^d)\asymp p\binom{n}{d} =\Omega(n\log n)$.
For the case $k<\delta n$, we see that 
\begin{align}
&|\mathcal{F}_{k}| \geq \binom{k}{1} \binom{n-k}{d-1}\geq k \binom{(1-\delta)n}{d-1} \nonumber \\
& \underset{n\to \infty}{\overset{(a)}{=}} (1+o(1))k (1-\delta)^{d-1}\binom{n-1}{d-1}\,, \label{lb1}
\end{align}
where ($a$) follows since
\begin{align}
\lim_{n\to \infty} \frac{\alpha^{d-1} \binom{n-1}{d-1}}{\binom{\alpha n}{d-1}}=1
\end{align} holds for a fixed $d$ and $\alpha\in(0,1)$. Hence, we get 
\begin{align}
& \sum_{k=1}^{\delta n} \binom{n}{k}(1-p')^{|\mathcal{F}_{k}|} \leq \sum_{k=1}^{\delta n} n^k e^{-(1+o(1))p'k (1-\delta)^{d-1} \binom{n}{d-1}} \nonumber\\
&= \sum_{k=1}^{\delta n}  e^{k\cdot \left\{\log n -(1+o(1))p'(1-\delta)^{d-1} \binom{n}{d-1} \right\}}\,. \label{ub4}
\end{align}
By choosing $\delta$ arbitrarily small, under the claimed condition, one can make
\begin{align*}
&p'(1-\delta)^{d-1} \binom{n}{d-1} = (1+o(1)) (1-\delta)^{d-1} \binom{n}{d}p'  \frac{d}{n} \\
&\geq  (1+\epsilon/2) \log n\,,
\end{align*}
which implies that \eqref{ub4} converges to $0$ as $n$ tends to infinity.
\subsection{Converse proof} As the parity measurement is symmetric,
\begin{align*}
\inf_{\psi}P_e(\psi) 
&=\Pr\left(\psi_{\text{ML}}(\mathbf{Y}) \notin \{\mathbf{X},~ \mathbf{X}\oplus \mathbf{1}\}~|~ \mathbf{X} = \mathbf{0}\right)\,.
\end{align*}
As before, we define the success event as: 
\begin{align}
S :=\bigcap_{\mathbf{V}\neq \mathbf{0}, \mathbf{1} } \left[{\sf d_H}(\mathbf{V})> {\sf d_H}(\mathbf{0})\right]\,. \label{def:s}
\end{align}
Again, it suffices to show that $\Pr(S)\to 0$, and to this end, we construct a subset of nodes  such that any two nodes in the subset do not share the same hyperedge. Unlike the previous case, the subset is now defined as:
\begin{align}
\mathcal{R}_{\text{big}}:=\left\{1,2,\cdots,r \right\} \label{def:rbig}
\end{align}
where $r=\lceil \frac{n}{\log^7 n} \rceil$, and we erase every node in $\mathcal{R}_{\text{big}}$ which shares hyperedges with other nodes in $\mathcal{R}_{\text{big}}$ to obtain $\mathcal{R}_{\text{res}}$. In view of Lemma~\ref{lem:ind}, we have $|\mathcal{R}_{\text{res}}| \geq (1-o(1))r$ almost surely; let $\Delta$ be such event. Conditioning on $\Delta$, we enumerate $r/2$ many elements of $\mathcal{R}_{\text{res}}$ by $b_1,\cdots, b_{r/2}$. As there are no hyperedges that connect two nodes in $\mathcal{R}_{\text{res}}$, the events $\{[{\sf d_H}( \mathbf{e}_{b_k}) >{\sf d_H}(\mathbf{0})]\}_{1\leq k \leq r/2 }$ are mutually independent conditioned on $\Delta$.
Hence, we get:
\begin{align}
\Pr(S) &\lesssim \Pr\left (S ~|~ \Delta \right)\nonumber\\
&\leq \Pr\left ( \bigcap_{k=1}^{r/2} \left[ {\sf d_H}(\mathbf{e}_{b_k}) >{\sf d_H}(\mathbf{0}) \right] ~\bigg|~ \Delta \right) \nonumber \\ 
&= \Pr\left (  {\sf d_H}( \mathbf{e}_{b_1}) >{\sf d_H}(\mathbf{0})  ~\big|~ \Delta \right)^{r/2}\,. \label{ubcon} 
\end{align}
Let $p' = (1-e^{-{\sf D_{KL}}(0.5\| \theta )})p$ as before. 
Using similar arguments used in the previous section, we have
	\begin{align}
	\Pr\left (  {\sf d_H}( \mathbf{e}_{b_1}) \leq {\sf d_H}(\mathbf{0})  ~\big|~ \Delta \right)\geq (1-o(1))e^{-p'\binom{n-1}{d-1}}\,. \label{keylb}
	\end{align}
This gives:
\begin{align*}
& \Pr\left (  {\sf d_H}( \mathbf{e}_{b_1}) >{\sf d_H}(\mathbf{0})  ~\big|~ \Delta \right)^{r/2}\\
&\leq \left(1-(1-o(1))e^{-p' \binom{n-1}{d-1}}\right)^{r/2}\\
&\leq  \exp\left(-(1-o(1))\frac{r}{2} \exp\left\{-p' \binom{n-1}{d-1} \right\} \right)\\
&\leq  \exp\left(-(1-o(1))\frac{n}{2\log ^7 n} \exp\left\{-(1+o(1))\cdot \frac{p' \binom{n}{d}d}{n} \right\} \right)\,.
\end{align*} 
Notice that the last term converges to $0$ as $\binom{n}{d}p'\leq (1-\epsilon)\frac{n\log n}{d}$, which completes the proof.

\section{Proof of Theorem~\ref{thm:main3}}
\label{pf:thm3}

When $d$ scales with $n$, a technical challenge arises, and we will focus on such technical difficulties, skipping most of the redundant parts.
\subsection{Proof of  the upper bound}
From \eqref{ub2} and \eqref{counting}, we get 
\begin{align}
P_e(\psi_{\text{ML}}) \leq \sum_{k=1} ^{n/2} \binom{n}{k}(1-p')^{N_k}\,, \label{exp:key}
\end{align}
where \begin{align}
	N_k:= \sum_{\substack{1 \leq i \leq d \\ i\text{ is odd}}}\binom{k}{i}\cdot\binom{n-k}{d-i} \label{exp:nk}
	\end{align}
	 and $p':= (\sqrt{1-\theta } -\sqrt{\theta})^2 p$.
Let us focus on counting $N_k$. When $d\asymp 1$, $\binom{n}{d} \approx \frac{n^d}{d!}$ suffices to obtain a proper bound on $N_k$.
However, in the general case where $d$ scales with $n$, one needs a more delicate bounding technique to obtain sharp results. 
The following lemma presents our new bound. 
\begin{lemma} ~\label{lemma:general}
	Let $\beta :=  \lceil \frac{n-d+1}{2d+1}  \rceil < n/2$ and $\alpha:=\frac{n-d+1}{d}$.
	Then 
	\begin{align*}
	\sum_{\substack{1 \leq i \leq d \\ i \text{ is odd}}} \binom{k}{i}\binom{n-k}{d-i}&\geq 
	\begin{cases}
	\frac{2k}{5\alpha}\binom{n}{d}, & \hbox{$k < \beta$;} \\
	\frac{1}{5}\binom{n}{d}, & \hbox{$\beta \leq k \leq n/2$\,.} 
	\end{cases}
	\end{align*}
\end{lemma}
\begin{IEEEproof}
	See Sec.~\ref{pf:tech}. The proof requires an involved combinatorial counting, which is one of our main technical contributions.
\end{IEEEproof}

Employing Lemma~\ref{lemma:general}, we get:
\begin{align}
\eqref{exp:key} \leq& \sum_{k=1}^{\beta-1} \binom{n}{k} (1-p')^{N_k} \nonumber + \sum_{k=\beta}^{n/2} \binom{n}{k} (1-p')^{N_k} \nonumber\\
\leq & \sum_{k=1}^{\beta-1} \binom{n}{k} (1-p')^{\frac{2k}{5\alpha}\binom{n}{d}}+\sum_{k=\beta}^{n/2} \binom{n}{k} (1-p')^{\frac{1}{5}\binom{n}{d}} \nonumber\\
\leq & \sum_{k=1}^{\beta-1} n^k e^{-p' \frac{2k}{5\alpha}\binom{n}{d}}+ 2^n e^{-\frac{1}{5}p'\binom{n}{d}} \nonumber\\
\leq & \sum_{k=1}^{\beta-1} \exp\left \{k\left(\log n- \frac{2p'\binom{n}{d}}{5\alpha}\right)\right \} \label{one}
\\ &+\exp \left \{n\log 2 -\frac{1}{5}p'\binom{n}{d} \right \}.\label{two} 
\end{align}
Note that \eqref{two} vanishes due to \eqref{ubb2}.
In order to show that \eqref{one} vanishes as well, we consider two cases: $d=o(n)$ and $d\asymp n$.
When $d=o(n)$,
\begin{align*}
&\sum_{k=1}^{\beta-1} \exp\left \{k\left(\log n- \frac{2p'\binom{n}{d}}{5\alpha}\right)\right \}	\\
\leq &\sum_{k=1}^{\beta-1} \exp\left \{k\left(\log n- \frac{2dp'\binom{n}{d}}{5n}\right)\right \} \\
\leq & \frac{\exp{\left(\log n - \frac{2dp'\binom{n}{d}}{5n}\right)}}{1 - \exp{\left(\log n - \frac{2dp'\binom{n}{d}}{5n}\right)}} \rightarrow 0,
\end{align*}
since $\log n - \frac{2dp'\binom{n}{d}}{5n} \rightarrow -\infty$.

If $d\asymp n$, 
	$$\sum_{k=1}^{\beta-1} \exp\left \{k\left(\log n- \frac{2p'\binom{n}{d}}{5\alpha}\right)\right \} \leq \beta \max_{1\leq k \leq \beta-1}{\exp\left \{k\left(\log n- \frac{2p'\binom{n}{d}}{5\alpha}\right)\right \} } = \beta \exp\left(\log n- \frac{2p'\binom{n}{d}}{5\alpha}\right),$$
	where the last equality holds since $\log n - \frac{2p'\binom{n}{d}}{5\alpha} < 0$, and hence $k=1$ achieves the maximum value. 
	Note that this vanishes since $\beta$ is asymptotically bounded by a constant.
Therefore, \eqref{one} always vanishes, completing the proof.

\subsection{Proof of the lower bound}

The lower bound statement can be rewritten as follows: $\inf_{\psi}P_e (\psi) \not\rightarrow 0$ if $\binom{n}{d} p \leq \max\left( (1-\epsilon) \frac{1}{d} \frac{n\log n }{(\sqrt{1-\theta}-\sqrt{\theta})^2}, \frac{n}{1-H(\theta)} \right)$.
	Note that when $d = \omega(\log n)$, the condition reduces to $\binom{n}{d} p \leq \frac{n}{1-H(\theta)}$.
	Hence, it is sufficient to show the following two statements.
	\begin{itemize}
		\item If $d = O(\log n)$: $\inf_{\psi}P_e (\psi) \not\rightarrow 0$ if $\binom{n}{d} p \leq \max\left( (1-\epsilon) \frac{1}{d} \frac{n\log n }{(\sqrt{1-\theta}-\sqrt{\theta})^2}, \frac{n}{1-H(\theta)} \right)$.
		\item If $d = \omega(\log n)$: $\inf_{\psi}P_e (\psi) \not\rightarrow 0$ if $\binom{n}{d} p \leq \frac{n}{1-H(\theta)}$.
	\end{itemize}
	
	We first show that $\binom{n}{d} p \leq \frac{n}{1-H(\theta)}$ implies $\inf_{\psi}P_e (\psi) \not\rightarrow 0$ for all $d$.
	By rearranging terms, we have $\binom{n}{d} p \leq \frac{n}{1-H(\theta)} \Leftrightarrow \frac{n}{\binom{n}{d} p} \geq 1-H(\theta)$. 
	One can immediately observe that this implies $\inf_{\psi}P_e (\psi) \not\rightarrow 0$ since $\frac{n}{\binom{n}{d} p}$ (which can be viewed as the rate of a code) cannot exceed the Shannon capacity of the channel $1 - H(\theta)$.
	
	We now prove that $\binom{n}{d} p \leq (1-\epsilon) \frac{1}{d} \frac{n\log n }{(\sqrt{1-\theta}-\sqrt{\theta})^2}$ implies $\inf_{\psi}P_e (\psi) \not\rightarrow 0$ if $d = O(\log n)$.
	Further, we will focus on the case of $\binom{n}{d}p \asymp \frac{n\log n}{d}$ since this is the regime where the largest amount of information is available.
Again, it is enough to show that $\Pr(S)\to 0$, where $S$ is defined as \eqref{def:s}. By defining $\mathcal{R}_{\text{big}}, \mathcal{R}_{\text{res}},\Delta$ and $b_1,\cdots, b_{r/2}$ as before, we again obtain~\eqref{ubcon}: 
\begin{align}
\Pr(S)\leq \Pr\left (  {\sf d_H}( \mathbf{e}_{b_1}) >{\sf d_H}(\mathbf{0})  ~\big|~ \Delta \right)^{r/2}\,.
\end{align}

We finish the proof by showing the following for the considered case:
\begin{align}
\Pr\left (  {\sf d_H}( \mathbf{e}_{b_1}) \leq {\sf d_H}(\mathbf{0})  ~\big|~ \Delta \right)\geq (1-o(1))e^{-2p'\binom{n-1}{d-1}}\,. \nonumber
\end{align}
While following the proof of \eqref{tight}, the key technical difficulty arises when checking $q=\Omega(\log n)$ (see \eqref{div}): a simple calculation yields $|\mathcal{F}_{b_1}|=\binom{n-1}{d-1}$ and $|\mathcal{G}_{b_1}|=\binom{n-|\mathcal{R}_{\text{big}}|}{d-1}$, but here it is not clear whether $\binom{n-|\mathcal{R}_{\text{big}}|}{d-1}\asymp \binom{n-1}{d-1}$ when $d$ is not a constant.
 We resolve this using a careful estimation as follows.
	As  $|\mathcal{R}_{\text{big}}|=\Theta(\frac{n}{\log^7 n})$ and $d=O(\log n)$, it is straightforward to verify
\begin{align*}
 1-\frac{1}{\log^2 n}\leq \frac{n-|\mathcal{R}_{\text{big}}|-j}{n-1-j}
\end{align*}
for $0\leq j \leq d-2$. 
This simple yet crucial inequality concludes: 
\begin{align*}
&\frac{\binom{n-|\mathcal{R}_{\text{big}}|}{d-1}}{\binom{n-1}{d-1}} = \prod_{j=0}^{d-2}\frac{n-|\mathcal{R}_{\text{big}}|-j}{n-1-j}\\
&\geq \left(1-\frac{1}{\log^2 n}\right)^{d-1}\approx \exp\left\{-\frac{d-1}{\log^2 n}\right\} \to 1.
\end{align*}

\subsection{Proof of Lemma~\ref{lemma:general}}
\label{pf:tech}
Without loss of generality, we prove the lemma assuming that $k \geq d$. 
The proof for the other cases is similar. 

We wish to obtain lower bounds on
\begin{align}
N_k=\sum_{\substack{1 \leq i \leq d \\ i\text{ is odd}}}\binom{k}{i}\binom{n-k}{d-i} = \underbrace{\binom{k}{1}\binom{n-k}{d-1}}_{\text{boundary odd term}}+ \underbrace{\sum_{i=1,3,\cdots,d-3,d-1} \binom{k}{i}\binom{n-k}{d-i}}_{\text{intermediate odd terms}} + \underbrace{\binom{k}{d-1}\binom{n-k}{1}}_{\text{boundary odd term}} \label{odd1}
\end{align}
in terms of $\binom{n}{d}$.
First, observe that
\begin{align}
\binom{n}{d} =\sum_{0\leq i\leq d} \binom{k}{i}\binom{n-k}{d-i}=\underbrace{\binom{k}{0}\binom{n-k}{d}}_{\text{boundary term}}+ \underbrace{\sum_{i=1,2,\cdots,d-2,d-1} \binom{k}{i}\binom{n-k}{d-i}}_{\text{intermediate terms}} + \underbrace{\binom{k}{d}\binom{n-k}{0}}_{\text{boundary term}}. \label{intact}
\end{align}
Suppose we have the following bounds:
\begin{align}
\underbrace{\binom{k}{0}\binom{n-k}{d} + \binom{k}{d}\binom{n-k}{0}}_{\text{sum of boundary terms}}&\leq A_1\underbrace{\left[ \binom{k}{1}\binom{n-k}{d-1} + \binom{k}{d-1}\binom{n-k}{1}  \right]}_{\text{sum of boundary odd terms}}; \label{bd:bdy}\\
\underbrace{\sum_{i=1,2,\cdots,d-2,d-1} \binom{k}{i}\binom{n-k}{d-i}}_\text{intermediate terms} &\leq A_2 \underbrace{\cdot \sum_{i=1,3,\cdots,d-3,d-1} \binom{k}{i}\binom{n-k}{d-i}}_\text{intermediate odd terms} + A_3N_k\,, \label{bd:int}
\end{align}
for some quantities $A_1,A_2, A_3>0$.
Then, by summing up the two inequalities, one can obtain a lower bound on $N_k$: 
\begin{align}\label{eq:alltogether}
\binom{n}{d} \leq \left(\max(A_1,A_2) + A_3\right) N_k\,.
\end{align}
Thus, the proof is completed as long as one can find the quantities $A_1, A_2$ and $A_3$ that satisfy \eqref{bd:bdy} and \eqref{bd:int}.

We begin with \eqref{bd:int}.
The following lemma asserts that $A_2 = 2$ and $A_3 = 3$ satisfy \eqref{bd:int}.
\begin{lemma}\label{lem:body} For $1 \leq k \leq n/2$, 
	$$\sum_{i=1,2,\cdots,d-2,d-1} \binom{k}{i}\binom{n-k}{d-i} \leq 2 \cdot \sum_{i=1,3,\cdots,d-3,d-1} \binom{k}{i}\binom{n-k}{d-i} + 3N_k.$$
\end{lemma}
\begin{IEEEproof}
	See Sec.~\ref{pf:body}.
\end{IEEEproof}

For \eqref{bd:int}, the following lemma characterizes $A_1$. 
\begin{lemma}\label{lem:tail1}
	Let $\beta := \left\lceil  \frac{n-d+1}{2d+1} \right\rceil$.
	For $\beta \leq k \leq n/2$, 
	\begin{align}
	\binom{k}{0}\binom{n-k}{d} + \binom{k}{d}\binom{n-k}{0} \leq 2\left[ \binom{k}{1}\binom{n-k}{d-1} + \binom{k}{d-1}\binom{n-k}{1}  \right].
	\end{align}
	For $k< \beta$, 
	\begin{align}
	\binom{k}{0}\binom{n-k}{d} + \binom{k}{d}\binom{n-k}{0} \leq \frac{\alpha}{k}\left[ \binom{k}{1}\binom{n-k}{d-1} + \binom{k}{d-1}\binom{n-k}{1}  \right]
	\end{align}
	and 
	\begin{align}
	\frac{\alpha}{k}\geq 2\,,
	\end{align}where $\alpha = \frac{n-d+1}{d}$.
\end{lemma}
\begin{IEEEproof}
	See Sec.~\ref{pf:tail1}.
\end{IEEEproof}
That is, $A_1 = 2$ if $\beta \leq k \leq n/2$, and $A_1 = \frac{\alpha}{k}$ if $k < \beta$.

We now are ready to prove Lemma~\ref{lemma:general} with the help of Lemma~\ref{lem:body}, Lemma,~\ref{lem:tail1} and \eqref{eq:alltogether}.
When $\beta \leq k<n/2$, $$\binom{n}{d} \leq 5 N_k.$$
When $k <\beta$, 
$$\binom{n}{d}\leq \left(\max\left(2, \frac{\alpha}{k}\right) + 3\right)N_k \leq \frac{5\alpha}{2k} N_k,$$
where the last inequality holds since $\frac{\alpha}{k} \geq 2$.
This completes the proof.

\section{Experimental results}
\label{sec:simulation}

\subsection{The homogeneity measurement case} \label{sec:exhom}
\subsubsection{Efficient algorithms} 
We also develop a computationally-efficient algorithm that achieves the information-theoretic limit characterized in Theorem~\ref{thm:main1}. Here we only present the algorithm while deferring a detailed analysis to our companion paper~\cite{ahn2017hypergraph}. The algorithm operates in two stages, beginning with a decent initial estimate from Hypergraph Spectral Clustering~\cite{ahn2017hypergraph} followed by iterative refinement. Detailed procedures are presented in Algorithm~\ref{alg}. 
Our algorithm is inspired by two-stage approaches that have been applied to a wide variety of problems including matrix completion~\cite{keshavan2010matrix,jain2013low}, phase retrieval~\cite{netrapalli2013phase,candes2015phase}, robust PCA~\cite{yi2016fast}, community recovery~\cite{abbe2016exact,abbe2015community,chen2016community,chin2015stochastic,gao2015achieving}, EM-algorithm~\cite{balakrishnan2017}, and rank aggregation~\cite{chen2015spectral}.

\begin{algorithm}
	\caption{}
	\label{alg}
	\begin{algorithmic}[1]
		\State For $E\in \binom{[n]}{d}$, define 
		\begin{align*}
			W_E := \begin{cases}
			 Y_E & \text{if } E\in\mathcal{E}; \\
			 0,  & \text{otherwise}.
			\end{cases}
		\end{align*}
		\State  Apply Spectral Hypergraph Clustering~\cite{ahn2017hypergraph} to a weighted hypergraph $([n], \{W_E\}_{E\in \binom{[n]}{d}})$ to obtain $\mathbf{X}^{(0)} =\{X^{(0)}_i\}_{1\leq i \leq n} \in \{0,1\}^n$.  
		\State For $t=0,1,\cdots, T-1$ ($T = c \log n$ for some constant $c>0$), update $\mathbf{X}^{(t)}= \{X^{(t)}_i\}_{1\leq i \leq n} $  as per
		\begin{align*}
			&X^{(t+1)}_i  = \begin{cases} X^{(t)}_i &\text{if}~{\sf d_H}(\mathbf{X}^{(t)}) <{\sf d_H}(\mathbf{X}^{(t)}\oplus \mathbf{e}_i);\\ X^{(t)}_i \oplus 1 &\text{if}~{\sf d_H}(\mathbf{X}^{(t)}) \geq {\sf d_H}(\mathbf{X}^{(t)}\oplus \mathbf{e}_i), \end{cases}		\end{align*}
		for $i=1,2,\cdots, n$, where ${\sf d_H}(\cdot)$ is defined in \eqref{defs}.
		\State Output $\mathbf{X}^{(T)}= \{X^{(T)}_i\}_{1\leq i \leq n}$.
	\end{algorithmic}
\end{algorithm}

\subsubsection{Performance of Algorithm~\ref{alg}}
We demonstrate the performance of Algorithm~\ref{alg} by running Monte Carlo simulations. 
\begin{figure}
	\centering
	\begin{subfigure}[b]{0.4\columnwidth}
		\includegraphics[width=\textwidth]{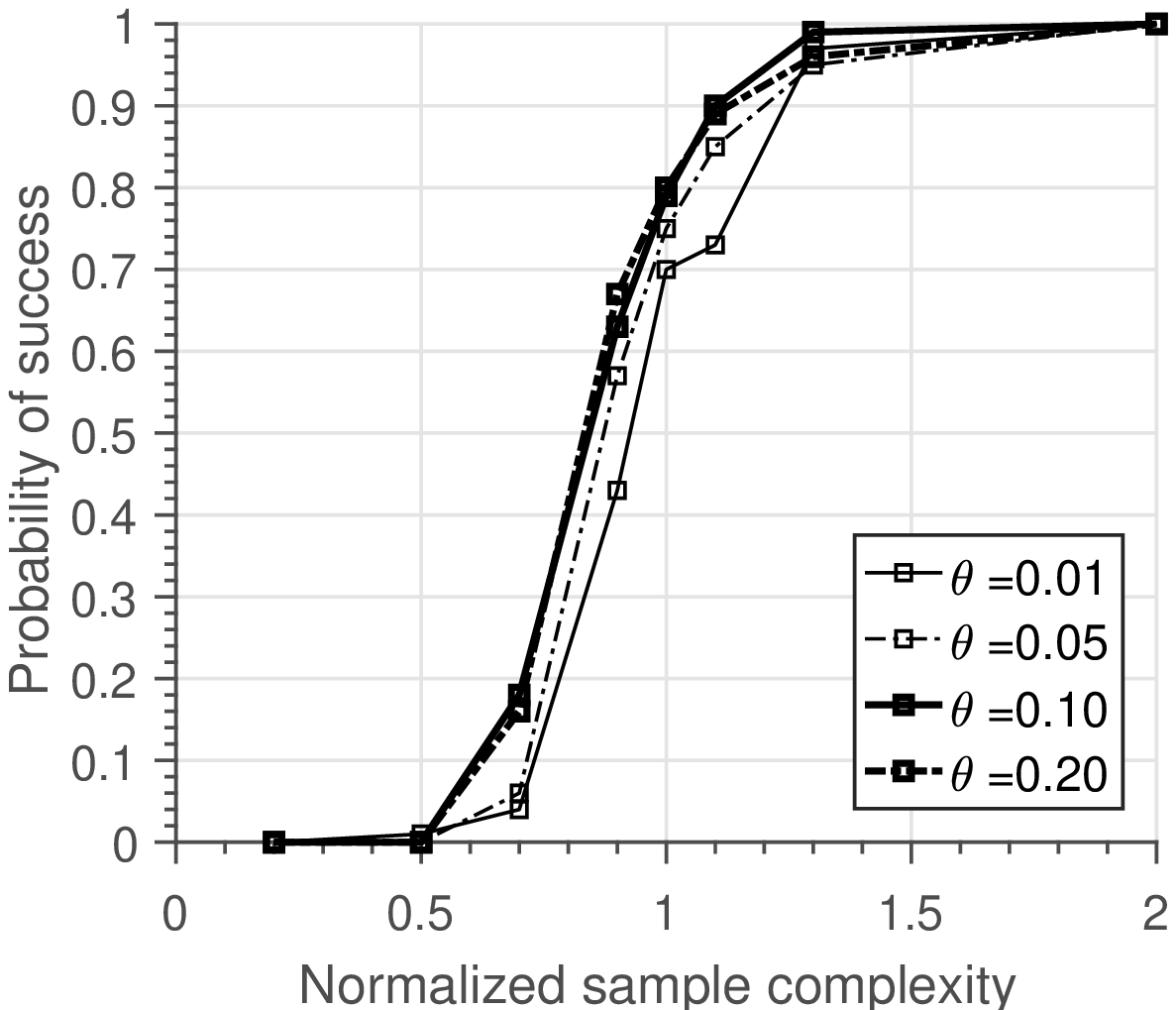}
		\caption{\footnotesize{Varying $\theta$}}
		\label{fig:3-a}
	\end{subfigure}	
	\begin{subfigure}[b]{0.4\columnwidth}
		\includegraphics[width=\textwidth]{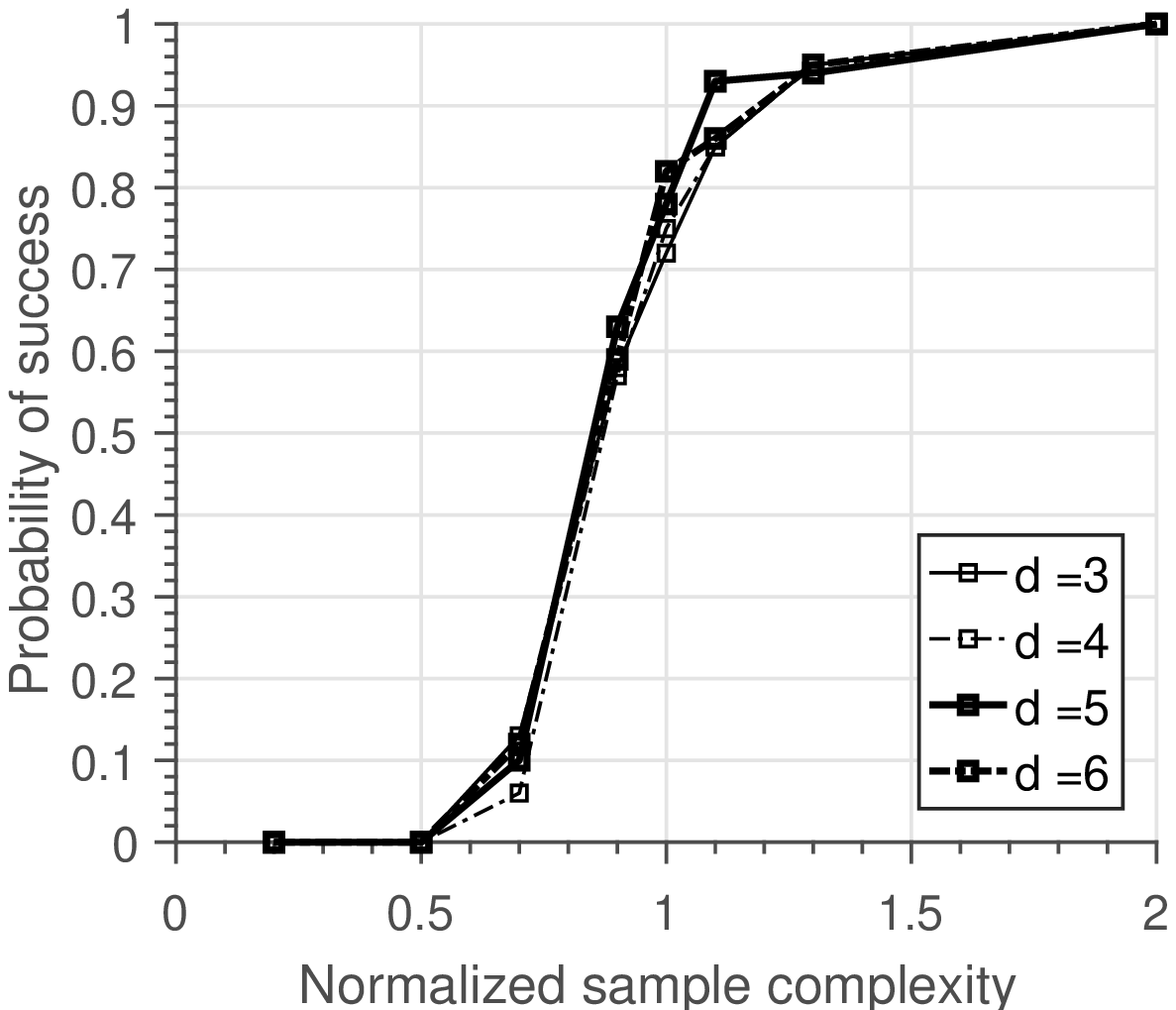}
		\caption{\footnotesize{Varying $d$}}
		\label{fig:3-b}
	\end{subfigure}	
	\caption{\footnotesize{\textbf {Algorithm~\ref{alg} achieves the optimal sample complexity.} We run Monte Carlo simulations to estimate the probability of success when: (a) $n=1000$, $d=4$, and for various choices of $\theta$; (b) $n=1000$, $\theta=0.05$, and for various choices of $d$. For each curve, we normalize the number of samples by the respective information theoretic limits, characterized in Theorem~\ref{thm:main1}. Observe that the probability of success quickly approaches $1$ as the normalized sample complexity crosses $1$.}}
	\label{fig:3}	
\end{figure}
Each point plotted in Fig.~\ref{fig:3-a} and Fig.~\ref{fig:3-b} indicates an empirical success rate. 
We take $100$ Monte Carlo trials. 
Fig.~\ref{fig:3-a} shows the probability of success when $n=1000$, $d=4$, and for various choices of $\theta$.
Shown in Fig.~\ref{fig:3-b} is the performance of our algorithm with $n=1000$, $\theta=0.05$, and for various choices of $d$.
For both figures, the $x$-axis denotes the number of samples normalized by the respective information-theoretic limits, characterized in Theorem~\ref{thm:main1}. 
One can observe that the success probability due to Algorithm~\ref{alg} quickly approaches $1$ as the normalized sample complexity crosses $1$, which corroborates our theoretical findings.

\begin{figure}
	\centering
	\begin{subfigure}[b]{0.4\columnwidth}
			\includegraphics[width=\textwidth]{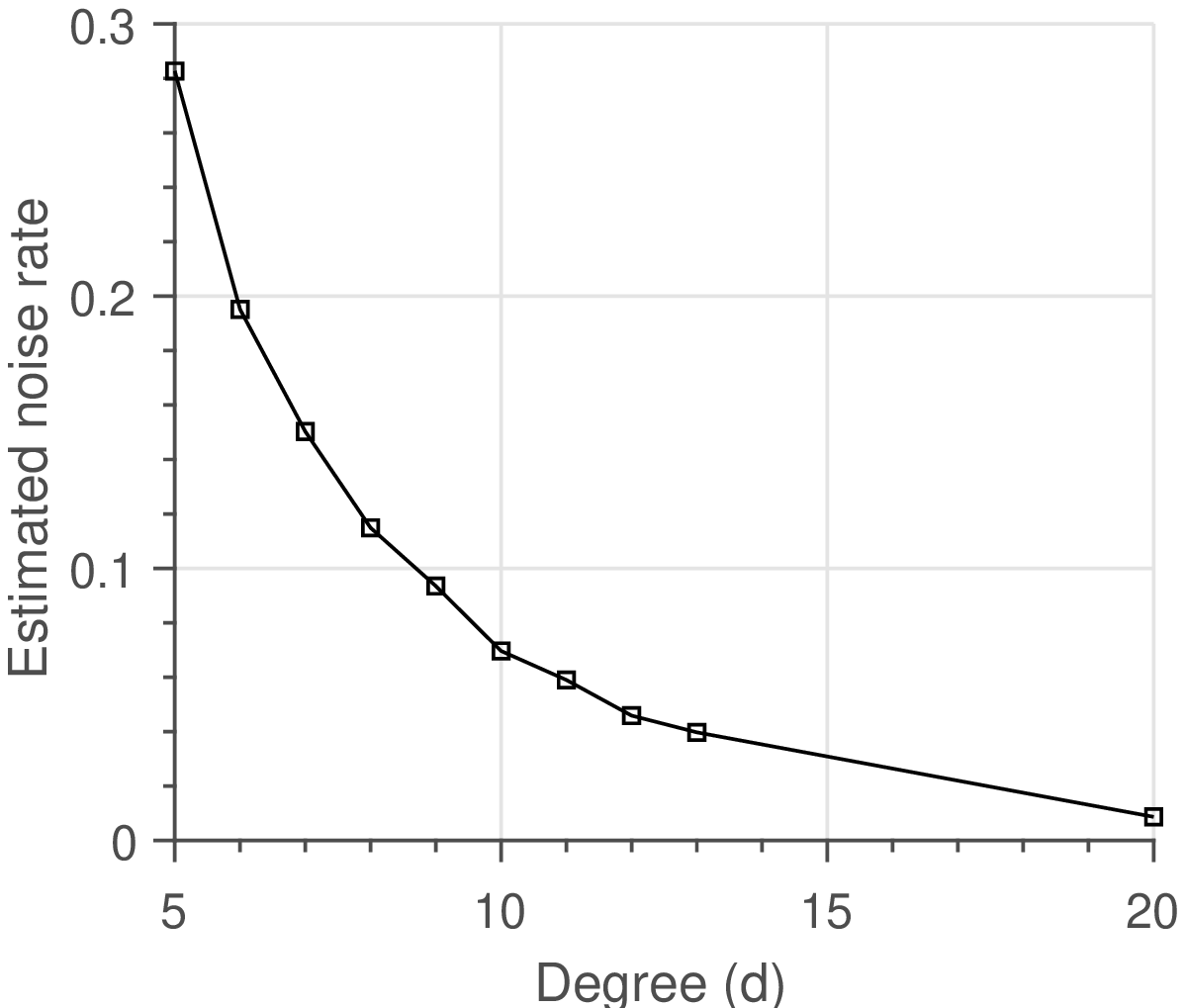}
		\caption{\footnotesize{Estimated empirical noise rate $\hat{\theta}$}}
		\label{fig:4-a}
	\end{subfigure}	
	\begin{subfigure}[b]{0.4\columnwidth}
		\includegraphics[width=\textwidth]{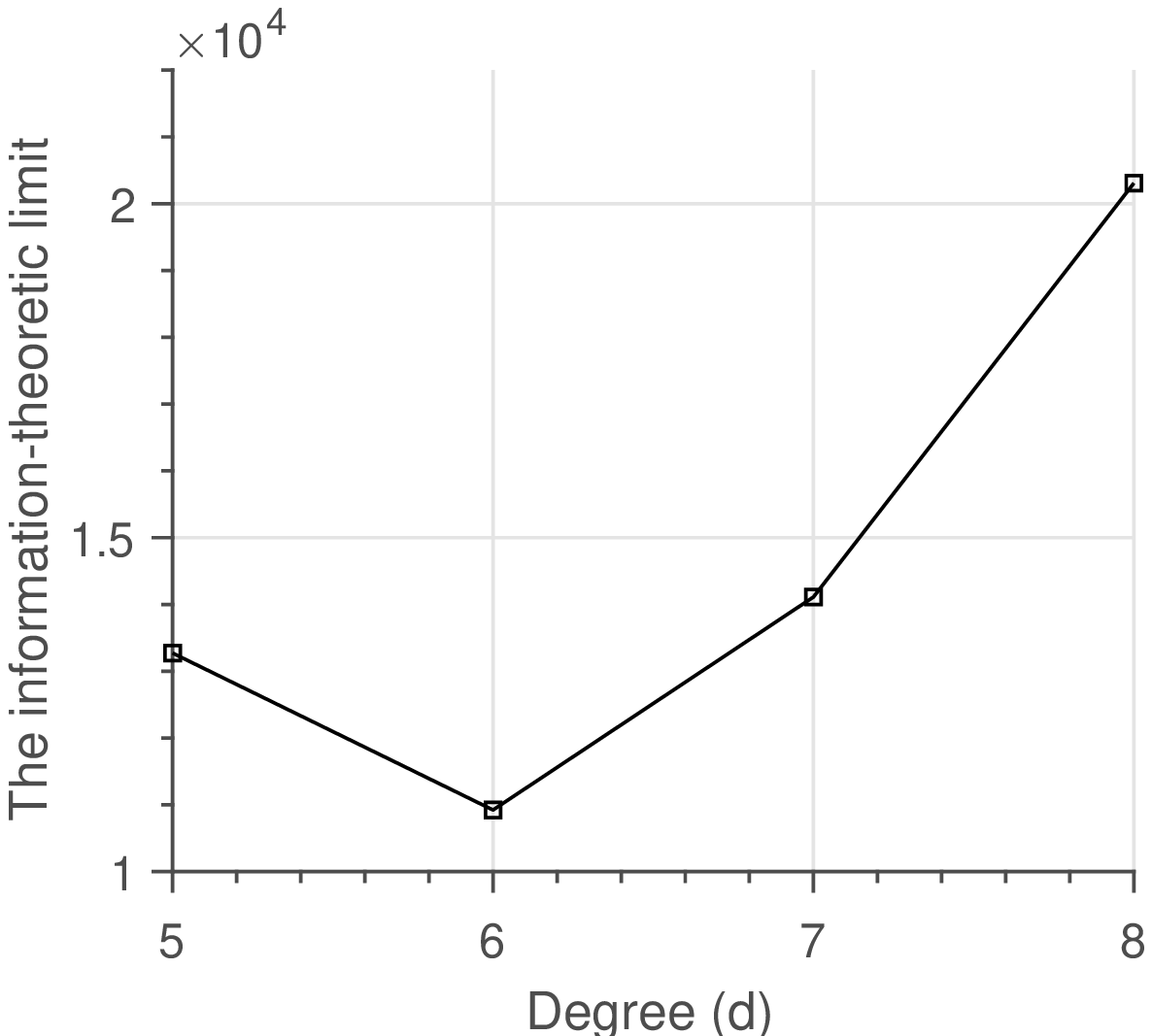}
		\caption{\footnotesize{$d^*$}}
		\label{fig:4-b}
	\end{subfigure}	
	
	\caption{\footnotesize{\textbf{Existence of $d^*$ in motion segmentation.} (a) We estimate the empirical noise rate $\hat{\theta}$ as a function of $d$ in motion segmentation. (b) We plug $\hat{\theta}$ to the limit characterized in Theorem~\ref{thm:main1} and verify that $d^*=6$.}}
	\label{fig:4}	
\end{figure}
\begin{figure}
	\centering
	\begin{subfigure}[b]{0.4\columnwidth}
		\includegraphics[width=\textwidth]{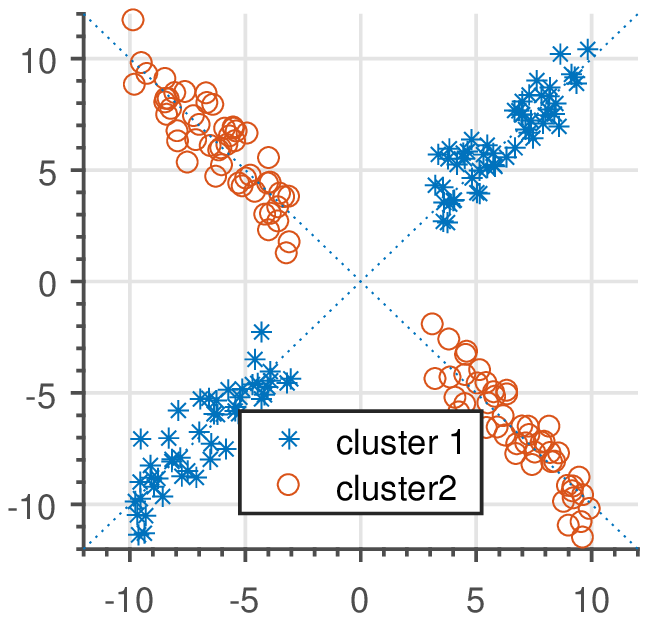}
		\caption{\footnotesize{Randomly generated data set}}
		\label{fig:5-a}
	\end{subfigure}	
	\begin{subfigure}[b]{0.4\columnwidth}
		\includegraphics[width=\textwidth]{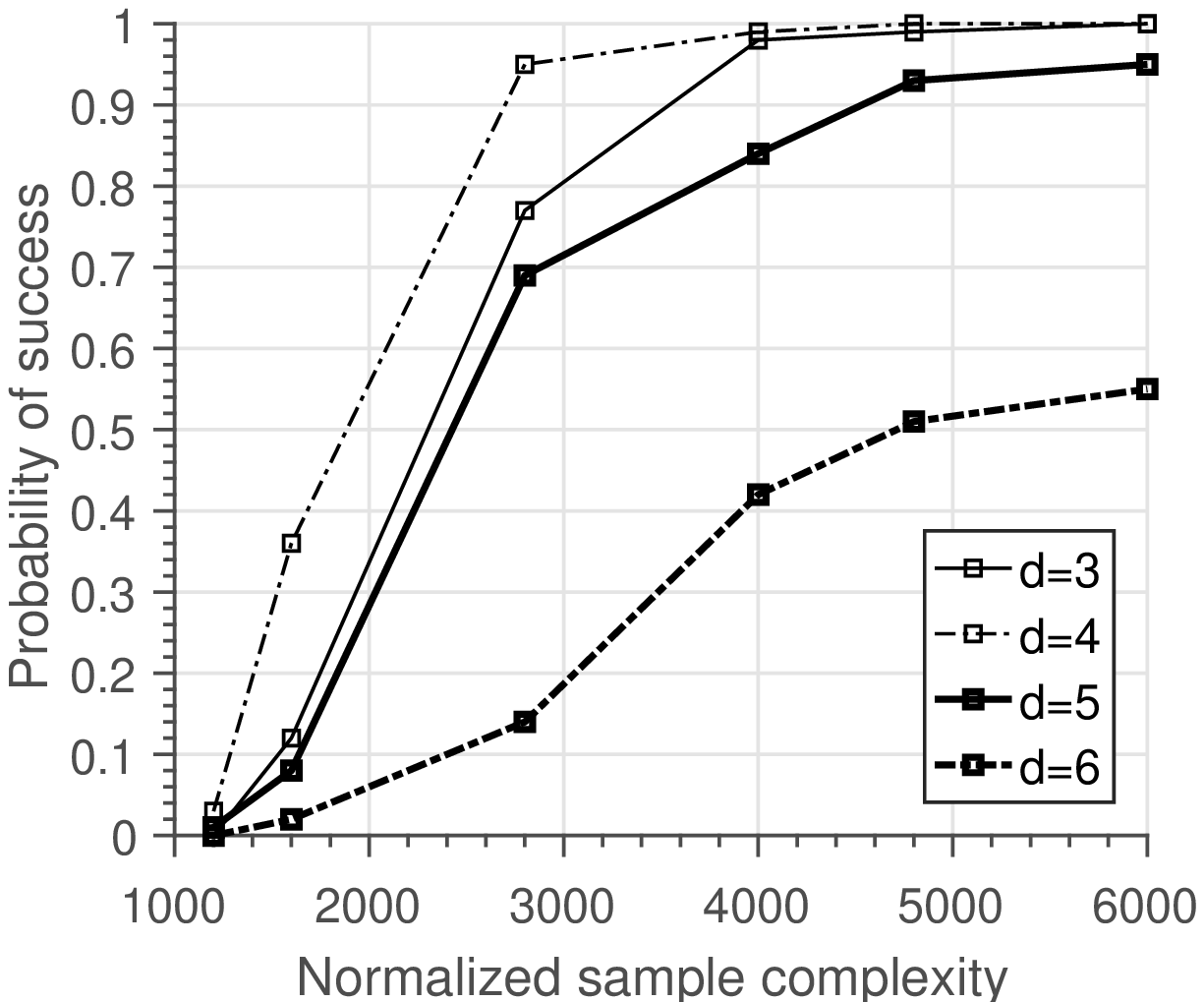}
		\caption{\footnotesize{Varying $d$}}
		\label{fig:5-b}
	\end{subfigure}	
	
	\caption{\footnotesize{\textbf{Optimal choice of $d$ when $\theta$ decays with $d$.} We run Monte Carlo simulations to estimate the probability of success with the data set shown in (a). We observe that the effective noise rate decreases as $d$ increases. For varying $d$ from $3$ to $6$, the success probability of Algorithm~\ref{alg} is shown in (b): the best performance of the algorithm is observed when $d=4$.}}
	\label{fig:5}	
\end{figure}

\subsubsection{Optimal $d$ for subspace clustering}
We observe how the fundamental limit varies as a function of $d$. 
As we briefly discussed in Sec.~\ref{sec:MainResults}, if the noise rate $\theta$ is irrelevant to $d$, the optimal choice of $d$ would be the minimum possible value of $d$.
However, if the noise quality $\theta$ depends on $d$, there may be a sweet spot for $d$.

We demonstrate the existence of a sweet spot in one of subspace clustering applications: motion segmentation. We use the benchmark Hopkins 155~\cite{tron2007benchmark} dataset to compute an empirical noise rate ${\theta}$ as a function $d$ as follows. 
For each sampled hyperedge $E=\{i_1,\cdots, i_d\}$, we adopt the method proposed in~\cite{chen2009spectral} to evaluate similarity between the corresponding $d$ data points that we denote by $D$. Then, we set $Y_E = 1$ if and only if $D$ is less than a fixed threshold, which is appropriately chosen so that $\Pr(Y_E=0 ~ |~ i_1,i_2,\cdots, i_d \text{ are from the same line})\approx \Pr(Y_E=1  ~|~ i_1,i_2,\cdots, i_d \text{ are not from the same line}).$
We estimate the effective noise rate $\hat{\theta} := \Pr(Y_E=0 ~ |~ i_1,i_2,\cdots, i_d \text{ are from the same line})$ for various $d$, and observe that $\hat{\theta}$ quickly decreases as $d$ increases; see Fig.~\ref{fig:4-a}.
We then plug these $\hat{\theta}$'s to the limit characterized in Theorem~\ref{thm:main1}; see Fig.~\ref{fig:4-b}. Note that $d=5$ is not the optimal choice, but $d=6$ is the sweet spot.

We also corroborate the existence of a sweet spot in a synthetic data set for subspace clustering, shown in Fig.~\ref{fig:5-a}. Here the goal is to cluster $n~(=200)$ $2$-dimensional data points approximately lying on a union of two lines ($1$-dimensional subspaces).
We compute $Y_E$ as above and evaluate the performance of Algorithm~\ref{alg}, shown in Fig.~\ref{fig:5-b}.
As a result, we observe that the optimal choice of $d$ here is $4$ rather than $3$. 

\subsection{The parity measurement case}\label{sim:parity}
\subsubsection{Efficient algorithms} \label{eff:parity}
For \pp~case, there are two efficient algorithms in the literature~\cite{watanabe2013message,jain2014provable}. In~\cite{watanabe2013message}, it is shown that for $d=3$, a variant of message passing algorithm successfully recovers the ground-truth vector provided that $\binom{n}{3}p =\Omega(n^2 /\log n)$.
Another efficient algorithm is based on a low-rank tensor factorization algorithm proposed in~\cite{jain2014provable}, and it is proved that reliable community recovery is feasible if $\binom{n}{3}p =\Omega(n^{1.5}\log^4 n)$. 
In either of the two cases, the sufficient condition comes with a polynomial term ($n$ or $n^{1/2}$) to the fundamental limit characterized in Theorem~\ref{thm:main1}.
In fact, it is conjectured in~\cite{florescu2015spectral} (see Conjecture 1 therein) that at least $n^{1.5}$ many samples are required for exact recovery.

On the other hand, focusing on the $\theta =0$ case, recovering the ground-truth vector from the measurement vector $\mathbf{Y}$ is essentially the same as solving linear equations over the Galois field of two elements $\mathbb{F}_2$.
Hence it immediately follows that efficient algorithms for solving linear equations such as Gaussian elimination can be employed in the noiseless case.

\begin{figure}
	\centering
	\includegraphics[width=.45\textwidth]{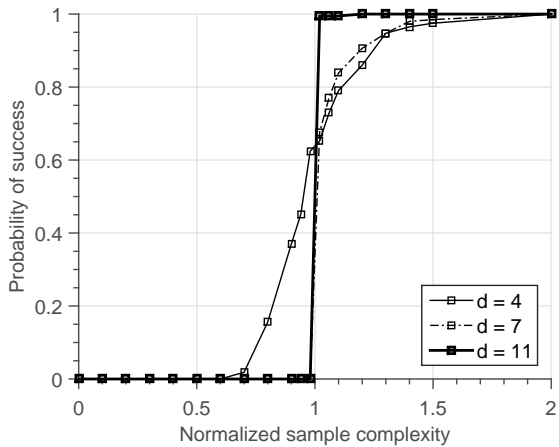}
	\caption{\footnotesize{We run the Monte Carlo simulations to estimate the probability of success for $n=1000$, varying $d$, and $\theta=0$. For each $d$, we normalize the number of samples by  $\max({n, n\log n/d})$. Observe that the probability of success quickly approaches $1$ as the normalized sample complexity crosses $1$.}
		\label{fig:6}}
\end{figure}
\begin{figure}
	\centering
	\includegraphics[width=0.45\textwidth]{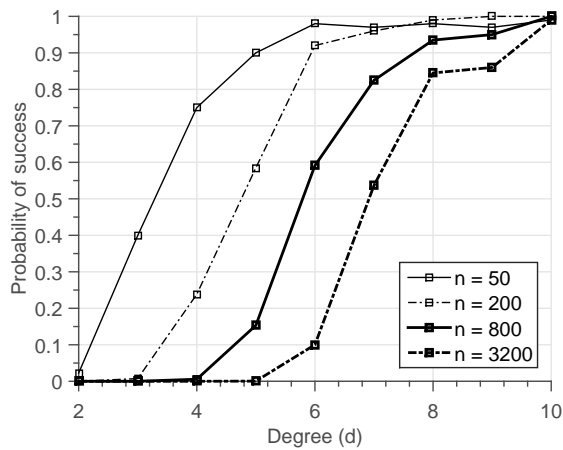}
	\caption{\footnotesize{We run the Monte Carlo simulations to estimate the probability of success for varying $n$, varying $d$, $\theta=0$, and $p = {1.1n}/{n \choose d}$.
			Note that when $n$ increases by a multiplicative factor of $4$, the curve shifts rightward about the same amount, supporting our result in Corollary~\ref{thm:main_dstar}}
		\label{fig:7}}
\end{figure}

\subsubsection{Information-theoretic limit} 
We first provide Monte Carlo simulation results which corroborate our theoretical findings in Theorem~\ref{thm:main2}.
Each point plotted in Fig.~\ref{fig:6} and Fig.~\ref{fig:7} is an empirical success rate. 
All results are obtained with $50$ Monte Carlo trials. 
In Fig.~\ref{fig:6}, we plot the probability of successful recovery for $n=1000$, varying $d$, and $\theta=0$. For each $d$, we normalize the number of samples by  $\max({n, n\log n/d})$. 
One can observe that the probability of success quickly approaches $1$ as the normalized sample complexity crosses $1$. 

\subsubsection{Minimum $d$ for linear sample complexity} Plotted in Fig.~\ref{fig:7} are the simulation results for varying $n$, varying $d$, $\theta=0$, and $p = {1.1n}/{n \choose d}$. 
We note that when $n$ increases by a multiplicative factor of $4$, the curve shifts rightward about the same amount, supporting our result in Corollary~\ref{thm:main_dstar}.

\section{Conclusion}
\label{sec:conclusion}

In this paper, we investigate the problem of community recovery in hypergraphs under the two generalized censored block models (GCBM), one based on \hh~and the other based on \pp. For these two models, we fully characterize the information-theoretic limits on sample complexity as a function of the number of nodes $n$, the size of edges $d$, the noise rate $\theta$, and the edge observation probability $p$. 
We also corroborate our theoretical findings via experiments.

We conclude our paper by highlighting  a few interesting open problems.
One interesting question is whether or not one can sharpen Theorem~\ref{thm:main3} to characterize exact information-theoretic limits for the scaling $d$ case. 
From the simulation results in Sec.~\ref{sim:parity}, we make the following conjecture: Under the setting of Theorem~\ref{thm:main3}, the information-theoretic limits is $\max\left\{ \frac{n}{1-H(\theta)},~ \frac{1}{d} \frac{n\log n}{(\sqrt{1-\theta}-\sqrt{\theta})^2} \right\}$.
Another interesting open problem is about the computational gap for \pp~case: Investigating efficient algorithms for this case would shed some light on the study of information-computation gaps.

\bibliographystyle{IEEEtran}
\bibliography{ref}

\appendices

\section{Proofs of Lemmas} \label{appenA}

\subsection{Proof of Lemma~\ref{lem:bound}}\label{app1}

Recall that 
\begin{align}
|\mathcal{F}_{i,j}| =&  \sum_{\ell=1}^{d-1} \binom{i}{\ell}\binom{k-i}{d-\ell} + \sum_{\ell=1}^{d-1} \binom{i}{\ell}\binom{n-k-j}{d-\ell}+\sum_{\ell=1}^{d-1} \binom{j}{\ell}\binom{n-k-j}{d-\ell}+ \sum_{\ell=1}^{d-1} \binom{k-i}{\ell}\binom{j}{d-\ell}\label{f2}.
\end{align}
In order to prove the lemma, it is sufficient to prove the following inequalities:
\begin{align}\label{eq:l1s1}
Z:= \sum_{\ell=1}^{d-1} \binom{i}{\ell}\binom{k-i}{d-\ell} + \sum_{\ell=1}^{d-1} \binom{i}{\ell}\binom{n-k-j}{d-\ell} \geq i\cdot\frac{(1-2\delta)^{d-1}}{2^{d-2}}  \binom{n-1}{d-1}
\end{align}
and 
\begin{align}\label{eq:l1s2}
\sum_{\ell=1}^{d-1} \binom{j}{\ell}\binom{n-k-j}{d-\ell}+ \sum_{\ell=1}^{d-1} \binom{k-i}{\ell}\binom{j}{d-\ell} \geq  j\cdot\frac{(1-2\delta)^{d-1}}{2^{d-2}}  \binom{n-1}{d-1}.
\end{align}
Here, we will focus on proving \eqref{eq:l1s1}.
We remark that the proof of \eqref{eq:l1s2} is essentially identical.

Since $i < \delta n$ and $j < \delta n$, 
\begin{align}
Z \geq \sum_{\ell=1}^{d-1} \binom{i}{\ell}\binom{k-\delta n}{d-\ell} + \sum_{\ell=1}^{d-1} \binom{i}{\ell}\binom{n-k-\delta n}{d-\ell}\label{f2}.
\end{align}
We further bound $Z$ by considering two cases separately: $k \geq \delta n$ and $k < \delta n$.
When $k \geq \delta n$,
\begin{align}
\sum_{\ell=1}^{d-1} \binom{i}{\ell}\binom{k-\delta n}{d-\ell} +\sum_{\ell=1}^{d-1} \binom{i}{\ell}\binom{n-k-\delta n}{d-\ell} &\geq i\binom{k-\delta n}{d-1} + i \binom{n-k-\delta n }{d-1} \nonumber\\
&\approx i\cdot \left[\left(\frac{k}{n}- \delta\right)^{d-1} + \left(1-\frac{k}{n}-\delta\right)^{d-1} \right]\binom{n-1}{d-1}\label{eq:60},
\end{align}
where the last inequality holds since $\binom{an}{b} \approx a^b\binom{n}{b} \approx a^b\binom{n-1}{b}$ for constants $a$ and $b$. 
We then apply H\"older's inequality: Given $p,q$ such that $1/p + 1/q = 1$, we have $\sum_{z}|x_z y_z| \leq \left( \sum_{z}|x_z|^p  \right)^{1/p} \left(  \sum_{z}|y_z|^q \right)^{1/q}$ for all sequences $\{x_z\}$ and $\{y_z\}$.
By setting $(x_1, x_2) = (\alpha, \beta), (y_1, y_2) = (1, 1), p = d-1, q = \frac{d-1}{d-2}$, we have
\begin{align}\label{eq:holder}
\alpha+\beta \leq (\alpha^{d-1}+\beta^{d-1})^{\frac{1}{d-1}} 2^{\frac{d-2}{d-1}}.
\end{align}
Applying \eqref{eq:holder} to \eqref{eq:60}, we have
\begin{align}\label{eq:l1r1}
i\cdot \left[\left(\frac{k}{n}- \delta\right)^{d-1} + \left(1-\frac{k}{n}-\delta\right)^{d-1} \right]\binom{n-1}{d-1} \geq i\cdot\frac{(1-2\delta)^{d-1}}{2^{d-2}}  \binom{n-1}{d-1}.	
\end{align}

When $k < \delta n$, $\sum_{\ell=1}^{d-1} \binom{i}{\ell}\binom{n-k-j}{d-\ell}$ becomes the dominant term. Hence,
\begin{align}\label{eq:l1r2}
\sum_{\ell=1}^{d-1} \binom{j}{\ell}\binom{n-k-\delta n}{d-\ell}+ \sum_{\ell=1}^{d-1} \binom{k-i}{\ell}\binom{\delta n}{d-\ell} &> i \binom{n-k-\delta n}{d-1} > i \binom{n-2\delta n}{d-1} \\
&\approx i \cdot (1-2\delta)^{d-1} \binom{n-1}{d-1} > i\cdot\frac{(1-2\delta)^{d-1}}{2^{d-2}}  \binom{n-1}{d-1}.
\end{align}
This completes the proof.

\subsection{Proof of Lemma~\ref{lem:ind}}
\label{pf:ind}
Denote by $\mathcal{R}_\text{big} \subset [n]$ the set of nodes of size $n/\log^7 n$. 
One can easily show that with high probability, some nodes of this set are connected by the same hyperedge(s).
Denote by $\mathcal{R}_\text{res}$ the largest subset of $\mathcal{R}_\text{big}$, whose elements do not share the same hyperedges.
The lemma states that with high probability, $|\mathcal{R}_\text{res}| \simeq |\mathcal{R}_\text{big}|$.

We now formally prove this statement.
Note that for a hyperedge $E = (i_1, i_2, \cdots, i_d)$, $|E \cap \mathcal{R}_\text{big}|$ is the number of nodes in $\mathcal{R}_\text{big}$ that are connected by the hyperedge. 
Hence, if $2 \leq |E \cap R_\text{big}| \leq d$, this hyperedge connects more than one nodes in $\mathcal{R}_\text{big}$, and $E \cap \mathcal{R}_\text{big}$ is the set of the nodes that share the same hyperedge $E$. 

Let us denote by $\mathcal{R}_\text{share}$ the subset of nodes that are connected by the same hyperedge(s).
Then, 
\begin{align}
\mathcal{R}_{\text{share}}:=  \bigcup_{k=2}^{d} \mathcal{R}_{\text{share}}^{(k)} := \bigcup_{k=2}^{d} \bigcup_{\substack{E\in \mathcal{E} : \\ |E\cap \mathcal{R}_{\text{big}}|=k }} E\cap \mathcal{R}_{\text{big}}.
\end{align}

Our proof strategy is as follows.
Since
\begin{align}
\mathcal{R}_{\text{res}}= \mathcal{R}_{\text{big}} - \mathcal{R}_{\text{share}} = \mathcal{R}_{\text{big}}-\bigcup_{k=2}^d \mathcal{R}_{\text{share}}^{(k)}, 
\end{align}
it is sufficient to show that 
\begin{align}\label{eq:lem3goal}
\left|\bigcup_{k=2}^d \mathcal{R}_{\text{share}}^{(k)}\right| = o(|\mathcal{R}_\text{big}|).
\end{align}
More specifically, we will show
\begin{align} \label{eq:lem3pbound}
\Pr \left( \exists k\in \{2,3,\ldots,d\}~\text{s.t.}~|\mathcal{R}^{(k)}_\text{share}| > \frac{n}{\log^9 n}\right)\rightarrow 0.
\end{align}
That is, with probability approaching $1$, $|\mathcal{R}_{\text{share}}^{(k)}| = o(n/\log^8 n)$ for all $k$, $2 \leq k \leq d$.
Note that this implies \eqref{eq:lem3goal} since
\begin{align}
\left|\bigcup_{k=2}^d \mathcal{R}_{\text{share}}^{(k)}\right| \leq \sum_{k=2}^d |\mathcal{R}_{\text{share}}^{(k)}| = O(d) \times o\left(\frac{n}{\log^8 n}\right) = o\left(\frac{n}{\log^7 n}\right) = o(|\mathcal{R}_\text{big}|).
\end{align}

In order to bound \eqref{eq:lem3pbound}, we first derive an upper bound on the expected value of $|\mathcal{R}^{(k)}_\text{share}|$.
By definition,
\begin{align}
|\mathcal{R}_\text{share}^{(k)}| \leq \sum_{\substack{E\in \mathcal{E} : \\ |E\cap \mathcal{R}_{\text{big}}|=k }} |E\cap \mathcal{R}_{\text{big}}| \leq \sum_{\substack{E\in \mathcal{E} : \\ |E\cap \mathcal{R}_{\text{big}}|=k }} k = |\{ E\in \mathcal{E} ~:~ |E\cap \mathcal{R}_{\text{big}}| =k\}|\cdot k.
\end{align}
Observe that $|\{ E\in \mathcal{E} ~:~ |E\cap \mathcal{R}_{\text{big}}|\}|$ is the sum of $\binom{|\mathcal{R}_{\text{big}}|}{k}\binom{n-|\mathcal{R}_{\text{big}}|}{d-k}$ i.i.d. Bernoulli random variables with probability $p$. 
Hence, 
\begin{align}
	\ex \left\{ |\{ E\in \mathcal{E} ~:~ |E\cap \mathcal{R}_{\text{big}}| =k\}|\cdot k \right\} & = k \binom{|\mathcal{R}_{\text{big}}|}{k}\binom{n-|\mathcal{R}_{\text{big}}|}{d-k}p \\ &= |\mathcal{R}_{\text{big}}| \binom{|\mathcal{R}_{\text{big}}|-1}{k-1} \binom{n-|\mathcal{R}_{\text{big}}|}{d-k}p\,. \label{eq:lem2_eq1}
\end{align}
As $|\mathcal{R}_\text{big}|=o(n)$, we have $\binom{|\mathcal{R}_{\text{big}}|-1}{k-1} \binom{n-|\mathcal{R}_{\text{big}}|}{d-k} \leq \binom{|\mathcal{R}_{\text{big}}|-1}{1} \binom{n-|\mathcal{R}_{\text{big}}|}{d-2}$, which in turn gives 
\begin{align*} 
\eqref{eq:lem2_eq1} &\leq |\mathcal{R}_{\text{big}}| \binom{|\mathcal{R}_{\text{big}}|-1}{1} \binom{n-|\mathcal{R}_{\text{big}}|}{d-2}p  \\
	&\leq 2|\mathcal{R}_{\text{big}}|^2 \binom{n-2}{d-2}p \\
	& = 2|\mathcal{R}_{\text{big}}|^2 \binom{n-2}{d-2}\frac{n \log n}{\binom{n}{d}} \\
	& = O\left(\frac{|\mathcal{R}_{\text{big}}|^2d^2\log n}{n}\right) = O\left(\frac{n}{\log^{11} n}\right),
\end{align*}
where the last equality holds since $\binom{n-2}{d-2} \approx  \binom{n}{d}\frac{d^2}{n^2}$.
Note that this inequality holds for any $2 \leq k \leq d$.
Using Markov's inequality,
\begin{align}
\Pr \left( |\{ E\in \mathcal{E} ~:~ |E\cap \mathcal{R}_{\text{big}}| =k\}|\cdot k > \frac{n}{\log^9 n}\right) &\leq \frac{\log^9 n}{n} \cdot O\left(\frac{n}{\log^{11} n}\right) = O\left(\frac{1}{\log^{2} n}\right).
\end{align}
Applying the union bound over all $2 \leq k \leq d$, 
\begin{align}
\Pr \left( \exists k\in \{2,3,\ldots,d\}~\text{s.t.}~|\{ E\in \mathcal{E} ~:~ |E\cap \mathcal{R}_{\text{big}}| =k\}|\cdot k > \frac{n}{\log^9 n}\right) \leq d \cdot O\left(\frac{1}{\log^{2} n}\right) = O\left(\frac{1}{\log n}\right).
\end{align}
This completes the proof.

\subsection{Proof of Lemma~\ref{lem:body}}\label{pf:body}

Since $1\leq i \leq \kjp-1$, 
\begin{align*}
&\frac{\binom{k}{i+1} \binom{n-k}{d-(i+1)}+\binom{k}{i-1}\binom{n-k}{d-(i-1)} }{ \binom{k}{i}\binom{n-k}{d-i}}\\
&= \frac{(k-i)(d-i)}{(i+1)(n-k-d+i+1)}+\frac{i(n-k-d+i)}{(k-i+1)(d-i+1)}\\
& \geq 2\sqrt{\frac{(k-i)(d-i)}{(i+1)(n-k-d+i+1)}\cdot \frac{i(n-k-d+i)}{(k-i+1)(d-i+1)}}\\
& =2\sqrt{\frac{(k-i)}{(k-i+1)}\cdot \frac{(d-i)}{(d-i+1)} \cdot \frac{i}{i+1}\cdot \frac{(n-k-d+i)}{(n-k-d+i+1)}}
\\ &\geq 2\sqrt{\left(\frac{1}{2}\right)^4} =\frac{1}{2}\,.
\end{align*}

\subsection{Proof of Lemma~\ref{lem:tail1}} \label{pf:tail1}
\begin{enumerate}
	\item $\beta\leq k \leq n/2$
	
	Since $d\leq n/2$ and $\beta < n/2$, one can verify the inequality using the following facts:
		\begin{align*}
		\binom{k}{0} \binom{n-k}{d}  \leq 2 \binom{k}{1}\binom{n-k}{d-1} &\Leftrightarrow k \geq \frac{n-d+1}{2d+1};\\
		\binom{k}{d}\binom{n-k}{0}\leq 2\binom{k}{d-1}\binom{n-k}{1} &\Leftrightarrow k \leq n - \frac{n-d+1}{2d+1}\,.
		\end{align*}
	\item $k<\beta$
	
	We first show that $\alpha / k \geq 2$.
	Since $k\leq \lceil\frac{n-d+1}{2d+1}\rceil-1 \leq \frac{n-d+1}{2d+1}$, 
	$\frac{\alpha}{k} =\left(\frac{n-d+1}{d}\right)/k > \left(\frac{n-d+1}{d}\right)/\left(\frac{n-d+1}{2d+1}\right) =\frac{2d+1}{d} \geq 2$. 
	Next, the inequality can be checked using the following facts:
	\begin{align*}
		\binom{k}{d}\binom{n-k}{0}\leq 2\binom{k}{d-1}\binom{n-k}{1} &\Leftrightarrow k \leq n - \frac{n-d+1}{2d+1}\\
	\frac{\binom{k}{0}\binom{n-k}{d}}{\binom{k}{1}\binom{n-k}{d-1}} = \frac{n-k-d+1}{kd}& \leq \frac{n-d+1}{kd} = \frac{\alpha}{k}\,.
	\end{align*}
\end{enumerate}

\end{document}